\documentclass[a4paper, english,american,12pt]{article}
\pdfoutput=1
\usepackage[T1]{fontenc}
\usepackage[latin9]{inputenc}
\usepackage{color}
\usepackage{babel}
\usepackage{array}
\usepackage{float}
\usepackage{rotfloat}
\usepackage{amsmath}
\usepackage{amssymb}
\usepackage{graphicx}
\usepackage{esint}
\makeatletter

\providecommand{\tabularnewline}{\\}

\@ifundefined{textcolor}{}
{%
 \definecolor{BLACK}{gray}{0}
 \definecolor{WHITE}{gray}{1}
 \definecolor{RED}{rgb}{1,0,0}
 \definecolor{GREEN}{rgb}{0,1,0}
 \definecolor{BLUE}{rgb}{0,0,1}
 \definecolor{CYAN}{cmyk}{1,0,0,0}
 \definecolor{MAGENTA}{cmyk}{0,1,0,0}
 \definecolor{YELLOW}{cmyk}{0,0,1,0}
}

\usepackage{babel}
\@ifundefined{definecolor}{color}{}
\usepackage{babel}
\@ifundefined{definecolor}{color}{}
\usepackage{babel}
\@ifundefined{definecolor}{color}{}
\usepackage{babel}
\usepackage{ifpdf}
\ifpdf 

 \IfFileExists{lmodern.sty}{\usepackage{lmodern}}{}

 \fi 

\let\myTOC\tableofcontents
\renewcommand{\tableofcontents}{%
  \frontmatter
  \pdfbookmark[1]{\contentsname}{}
  \myTOC
  \mainmatter }

\@ifundefined{showcaptionsetup}{}{%
 \PassOptionsToPackage{caption=false}{subfig}}
\usepackage{subfig}
\makeatother

\begin{document}
\global\long\def\Forcefriction{\overrightarrow{f_d}}

\global\long\def\ForcePropulsion{\overrightarrow{f_p}}

\global\long\def\ForceLennardJonnes{\overrightarrow{f_{LJ}}}

\global\long\def\Speedbacteria{v}

\global\long\def\Speedfluid{\overrightarrow{u}}

\global\long\def\SpeedTotal{U}

\global\long\def\ProbabilityTumble{P_{tumb}}

\global\long\def\ProbaDensTumbPop{\mathcal{P}_{tumb}}

\global\long\def\ProbabilityPhenotype{P_{ph}}

\global\long\def\ProbabilityDivision{P_{d}}

\global\long\def\MemoryShort{m_{s}}

\global\long\def\MemoryLong{m_{l}}

\global\long\def\TimeMemoryShort{\tau_{s}}

\global\long\def\TimeMemoryLong{\tau_{l}}

\global\long\def\CoefAer{\alpha}

\global\long\def\CoefRenorm{s}

\global\long\def\TimeAjustable{\xi}

\global\long\def\TimePheno{\tau_{ph}}

\global\long\def\TimeMatrix{\tau_{m}}

\global\long\def\TimeRun{\tau_{run}}

\global\long\def\TimeDiv{\tau_{d}}

\global\long\def\CoupureVol{K_{g}}

\global\long\def\Oxygen{c}

\global\long\def\BacterialCon{n}

\global\long\def\Radius{r}

\global\long\def\RadiusDiv{r_{d}}

\global\long\def\EpaisseurCouche{r}

\global\long\def\TimeStep{\mbox{\ensuremath{\Delta}t}}

\global\long\def\DistanceEntreParticle{d}

\global\long\def\VolumeMax{V_{m}}

\global\long\def\VolumeDiv{V_{d}}

\global\long\def\OxygenGradient{G_{0}}

\global\long\def\ProbabilityDensityAngle{\mathcal{N}_{\theta}}

\title{An individual-based model for biofilm formation at liquid surfaces}

\author{Maxime Ardré$^{+,*,o}$, Hervé Henry$^{+}$, Carine Douarche$^{*}$, Mathis Plapp$^{+}$\\
$^+$ Laboratoire PMC, Ecole Polytechnique CNRS,  Palaiseau, France\\
$^*$ Laboratoire de Physique des Solides, Universit\'e Paris-Sud CNRS, Orsay, France\\
$^o$ Laboratoire de Physique Statistique, ENS CNRS, Paris, France}
\maketitle

\begin{abstract}
The bacterium {\em Bacilus subtilis} frequently forms biofilms
at the interface between the culture medium and the air. We develop
a mathematical model that couples a description of bacteria as 
individual discrete objects to the standard advection-diffusion
equations for the environment. The model takes into account 
two different bacterial phenotypes. In the motile state, 
bacteria swim and perform a run-and-tumble motion that is biased 
toward regions of high oxygen concentration (aerotaxis). In the 
matrix-producer state they excrete extracellular polymers, 
which allows them to connect to other bacteria and to form a biofilm.
Bacteria are also advected by the fluid, and can trigger 
bioconvection. Numerical simulations of the model reproduce 
all the stages of biofilm formation observed in laboratory
experiments. Finally, we study the influence of various model 
parameters on the dynamics and morphology of biofilms. 
\end{abstract}

\section{Introduction}

Bacteria are unicellular prokaryotic microorganisms. They are ubiquitous
and constitute a large part of the terrestrial biomass. Bacteria can
live as individual cells during
the planktonic phase. However, most of the time, they are part of 
self-organized communities of complex architecture adsorbed on interfaces: 
the biofilms. Besides the bacteria themselves, biofilms are mostly
made of an extracellular matrix composed of macromolecules \cite{Branda2005,Branda2006,Pamp}
that are produced by the bacteria and lead to cohesive interactions between
them \cite{Costerton1995,Watnick2000}. Most of the time biofilms
appear in aqueous environments either on solid surfaces or at the
water-air interfaces. Due to this multicellular organization, the
bacteria have different properties than in the motile state. For
example, bacteria trapped in biofilms can exhibit an increased resistance
to antibiotics as well as to environmental stresses (desiccation,
UV radiation, disinfecting agents, shear flow...) \cite{Hoyle1991,Finlay1997}.
Therefore, the association in biofilms is a crucial step both for
survival and spreading of bacterial colonies \cite{OToole2000}.

Many environmental and genetic factors influence the development of
biofilms \cite{OToole2000}. Although biofilm formation is not
understood in all details,
a consistent picture of biofilm growth on solid substrates has been
proposed: bacteria in the planktonic phase anchor preferentially on
a stable surface, which initiates the nucleation of bacterial microcolonies
\cite{Watnick2000,OToole2000}. Bacteria constituting the microcolonies
secrete an extracellular matrix in which they embed, and form a mature
biofilm with a complex multiscale architecture \cite{Bridier2011,Vlamakis2008}.
Later on, bacteria can also detach from the superficial layers, return to
the planktonic state and spread to new parts of the surface\cite{Picioreanu2007}.

Much less is known on biofilm formation at water-air interfaces. We
have performed laboratory experiments on \textit{Bacillus subtilis} (\textit{B. subtilis})
and built a mathematical model that can reproduce the main experimental
observations. The details of the experimental results will be published
elsewhere; here, we briefly present a summary to motivate the construction
of the mathematical model.

\begin{figure}[htbp]
\centerline{\includegraphics[width=1\textwidth]{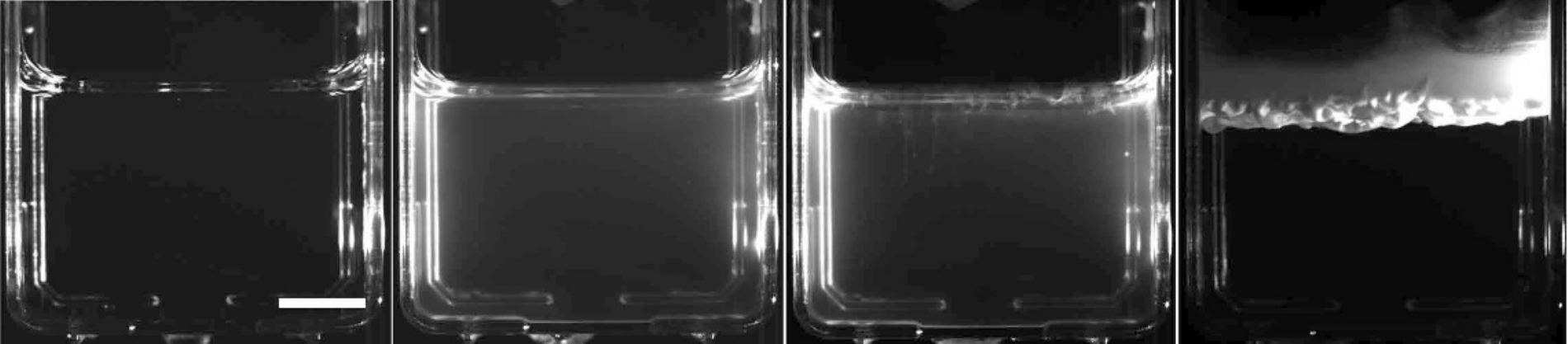}} \protect\caption{Side view of \textit{B. subtilis} pellicle developing on top of a
liquid culture media at T=$31^{\circ}$C at : (a) 0 hours, (b) 15
hours (c) 19 hours 30 minutes, and (d) 67 hours after incubation.
Images are acquired with a black and white CCD camera. The scale bar
is 1 cm long.}

\label{fig:experience} 
\end{figure}

\textit{Bacillus subtilis} is a stricly aerobic bacterium which is able to
form floating pellicles that can reach a thickness of several hundred
microns \cite{Branda2001} on the top of nutritive media. During a
typical experiment, the nutritive medium (Luria broth supplemented
with glycerol and $\textrm{MnO}_{2}$) is initially inoculated with
a small concentration of bacteria. Then, the cells divide and grow
over several hours to yield a homogeneous suspension (see Fig. \ref{fig:experience}
(a)). Later (Fig.~\ref{fig:experience} (b)), bacteria start to accumulate
close to the interface, and a rapid transition occurs (between (b)
and (c)), which leads to the nucleation of bacterial clusters on the
interface (Fig.~\ref{fig:experience} (c)). At this stage, some macroscopic  filaments,
which can be seen inside the liquid, are advected by fluid motion.
With time, the biofilm further develops into a mature floating pellicle
that exhibits a typical wrinkled morphology (Fig. \ref{fig:experience}(d)).

These experiments indicate several important phenomena that need
be included in the model. First, \textit{B. subtilis} is known
to exhibit aerotaxis \cite{Wong1995}, that is, to migrate to areas
in which the concentration of oxygen is high. In the experiments, the
water-air interface acts as an oxygen source; therefore, aerotaxis
is a possible explanation for the accumulation of bacteria close to the
surface. Second, planktonic bacteria are slightly more dense than
water. As a result, an accumulation of bacteria at the surface is
unstable and should drive the development of bioconvection \cite{Kessler1998},
which may hinder or help the biofilm formation. Finally,
there is a well-defined change in bacterium phenotype from motile
(planktonic phase) to matrix-producing (biofilm phase). This change
is controlled by a genetic switch that is believed to be triggered
by a \textit{quorum sensing} mechanism \cite{Kobayashi2007,Vlamakis2008}
(i.e. when the local density of bacteria exceeds a threshold value,
the phenotype changes).

Many different modelling strategies have already been proposed and
used for biofilm formation on solid substrates. It is possible to
average the contribution of individual bacteria and to write a full
continuum model \cite{Lindley2012,Pedley1992}. It is also possible
to consider the bacteria as individual objects in cellular automata
models (CA) \cite{Picioreanu1998} or individual-based models (IdbM)
\cite{Kreft2001}. The latter model can be hybridized with a continuum
model to describe the contribution of the environment. Modeling is
often accompanied by numerical simulations to study the processes
which structure the biofilm \cite{Xavier2005,Alpkvist2006,Nadell2013}.

Here, we formulate a detailed model for biofilm formation at liquid-air
interfaces. The main point is that the model captures the transition
between a ``gas'' of individual swimming bacteria and the biofilm
pellicle (a soft solid). Since this transition involves a change in
the connectivity between bacteria, a description using a full continuum
model is difficult. We have chosen a hybrid approach, in
which the bacteria are described as individual particles, whereas
the local environment (oxygen concentration, fluid velocity) is described
by continuous fields. In order to keep the model minimal, we only
consider two bacterial phenotypes: in the motile state, bacteria
are self-propelled swimmers that perform a standard ``run-and-tumble''
motion. The bacteria interact through a local repulsive potential to describe
collisions between bacteria (hydrodynamic interactions are neglected). 
In the matrix-producer
state, the bacteria stop moving actively, and produce macro-molecules
that constitute the extracellular matrix. Due to the presence of this
matrix, they are able to ``connect'' to other bacteria. In this
case, the interaction between bacteria is described by an attractive
potential. The transition between the two phenotypes is triggered
by a quorum-sensing mechanism.

This model has been implemented in two dimensions, using the 
\emph{discrete-element method} \cite{Lazarevic2002} to calculate 
the evolution of the bacteria, and the \emph{finite-difference method} 
\cite{FuidDyn} for the continuous fields, and simulations involving 
up to 10000 bacteria were performed.
This setting is sufficient to demonstrate that all the steps
of biofilm formation that are observed in the experiments can be
reproduced. The influence of the model parameters on these different
stages was also studied, yielding suggestions for further model 
improvements and experiments.

In the following, an overview of the model architecture is first given,
followed by a detailed description of each ingredient (section \ref{sec_model}).
In section \ref{sec_params}, the choice of the model parameters is
discussed in detail. In Section \ref{sec_results}, simulation results
are presented, which demonstrate the ability of the model to properly
describe biofilm formation. Finally, section \ref{sec_conclusion}
discusses the main conclusions and perspectives of this work.

\section{Model description}

\label{sec_model}

\subsection{Overview}

Our goal is to construct a minimal
mathematical model that reproduces the different steps of biofilm
formation, from motile bacteria (individuals swimming in the liquid)
all the way to the mature biofilm (bacteria linked by extracellular
matrix). In order to include the phenomena of aerotaxis and
bioconvection, the local oxygen concentration $\Oxygen$ and the 
fluid velocity $\Speedfluid$ are described as continuous 
fields which obey partial differential equations (PDE).

Bacteria are represented as discrete objects, with interactions that
depend on their internal state. Each bacterium is characterized by
its position, its velocity, and internal variables that reproduce
its behavior (aerotaxis, phenotype, cell cycle). One of these 
internal variables is the bacterial phenotype. We take into account
only two of them : \textit{motile} and \textit{matrix producer}.
In the motile phenotype, the bacteria propel themselves with a constant
velocity and change their direction with a frequency that is determined
by the local concentration in oxygen (\emph{run-and-tumble} motion).
Each motile bacterium increases its body size with time and divides
into two cells with a constant rate. In the matrix-producer state,
propulsion is absent, and the bacterium produces extracellular matrix,
which makes its volume grow with time (without division). Moreover,
we assume that the transition from the motile to the matrix-producer
phenotype is irreversible and triggered by a \textit{quorum sensing}
mechanism: bacteria tend to switch to the matrix-producer type when
a certain bacterial concentration is (locally) exceeded.

The main effect that is caused by changing the phenotype is to change
the interactions between bacteria. In both states, there is a repulsive
interaction that prevents bacteria from overlapping. When a bacterium has
started to produce matrix, which consists of ``sticky'' macromolecules,
it can establish \textit{links} with  bacteria when  they are  in contact.
These links correspond to a spring-like attractive interaction potential,
and break when the distance between the two bacteria is above a threshold
value.

In the following, we first describe how the discrete and continuum
approaches are coupled, and then give more details about the description
of the bacteria as discrete objects.

\subsection{Environment}

The geometry of the simulation setup is inspired by the experiments
depicted in Fig.~\ref{fig:experience} (a container that is open
at the top is filled with nutritious medium). { We restrict our
simulations to two dimensions, that is, the simulation domain
is a vertical plane. In order to convert two-dimensional densities
to three-dimensional ones that can be compared to values measured
in experiments, we assume a thickness of the sample of 10$\mu$mX2x4cm$^{2}$. 
While the exact value is arbitrary, it approximately corresponds 
to the average diameter of the bacteria in the model}. The top surface is
assumed to remain perfectly flat, and it is in contact with air. 
The domain is discretized using a regular square grid of $N\times N/2$
points, where $N=L/\Delta x$ with $L$ being the horizontal system size
(the height of the fluid layer is $L/2$), and $\Delta x$ the
grid spacing, with grid points being located on the walls
and on the fluid surface. 

Being interested in the continuum fields on macroscopic length
scales, we choose the grid spacing $\Delta x$ to be of the order of
a millimeter. On this scale, the bacteria (of micron size) are point-like
objects. Therefore, when an information about the environment of
a bacterium is needed, the value of the relevant variable is calculated using 
a bilinear interpolation of the values at the three closest grid points.
Conversely, the terms involving bacteria in the PDEs are computed
by averaging the contribution of neighboring bacteria, as described
below.

\subsubsection{Density fields}

\label{sec_density} In the continuum equations that are presented
below, two source terms are calculated from the positions of the 
individual bacteria: the number density for the total oxygen consumption,
and the mass density for the buoyancy force, as follows:
the neighborhood of a grid point (called ``cell'' in the following)
is defined as its Voronoi cell (the set of all points in space that 
are closer to it than to any other grid point). The number 
density $\BacterialCon$ is then defined by the number of bacteria
that are contained in the cell, divided by the cell volume (which is smaller
in the case of points located on the boundaries of the system). In
order to obtain a number density $\BacterialCon$ with units of m$^{-3}$
that can be compared to experimental measurements, we set the thickness
of the system to 10 $\mu$m, which is comparable to the  the size of a bacterium.

The local mass density is also dependent on the bacterial state. Various
values for the mass density of motile bacteria have been published
in the literature \cite{Pedley1992,Kessler1998,Cisneros2008these}.
Our experimental observations clearly indicate that the density of
motile bacteria is larger than that of water (they sediment in the absence of
active motion). The situation is not as clear for the matrix producers.
If the mature biofilm is cut into pieces, some of them float, while
others sink. This indicates that the biofilm density is, on average,
very close to that of the medium. Therefore, we assume that matrix-producing
bacteria have the same density as the medium, $\rho_{0}$, while motile
bacterial density $\rho_{b}$ is slightly larger. The local mass density
is then
\begin{equation}
\rho=\rho_{0}+\frac{V_{b}}{V_{{\rm cell}}}(\rho_{b}-\rho_{0}),\label{eq:density}
\end{equation}
where $V_{{\rm cell}}$ denotes the volume of the grid cell, and $V_{b}$
the total volume occupied by motile bacteria in the cell.

While this definition is straightforward, the fact that density fields
are dependent on the positions of the individual bacteria (which perform
a random walk) implies that they fluctuate in space and time. The
relative magnitude of these statistical fluctuations depends on the
average number of bacteria in a cell, and thus decreases when the
cell size is increased. The choice of the grid spacing is therefore
dictated by a compromise between the spatial resolution (which requires
small grid spacing) and the smoothing of the density field (which
requires large grid spacings). Typically, in our simulations we have
$N=20$ and up to 10000 bacteria, which yields around $50$ bacteria
per cell for a homogeneous system. As a result, typical relative fluctuations
are smaller than $0.2$, which is small enough to avoid spurious effects
on the computation of the fluid motion and on the evaluation of the
transition rate from motile to matrix producer phenotype.

\subsubsection{Fluid flow calculation }

Since the mass density of the bacteria is slightly higher than that of water, an accumulation of bacteria under the surface creates
an unstable stratification that can give rise to convection. The medium
is modeled as an incompressible Newtonian fluid with a mass density
that depends on the bacterial concentration {as it has been 
done previously in the work of Hillesdon \textit{et al} 
\cite{Hillesdon95,Hillesdon96}}. Its dynamics is described
by the Navier-Stokes (NS) equation in the Boussinesq approximation,
in which the variations of density are neglected except in the buoyancy
force. Since our system is two-dimensional, we use a vorticity/stream
function formulation which guarantees incompressibility, independently
of the magnitude of the numerical error. In this approach, the velocity
$\Speedfluid$ of the fluid derives from the stream function $\psi$,
\begin{equation}
u_{x}=\frac{\partial\psi}{\partial y}\qquad u_{y}=-\frac{\partial\psi}{\partial x}\,.\label{eq:ux}
\end{equation}
which is solution of 
\begin{equation}
\Delta\psi=-\omega\,,\label{eq:streamFunction}
\end{equation}
where the time evolution of the vorticity $\omega$ writes 
\begin{equation}
\frac{\partial\omega}{\partial t}+\left(\Speedfluid\cdot\overrightarrow{\nabla}\right)\omega=\nu\triangle\omega-g\frac{\partial}{\partial x}\frac{\rho}{\rho_{0}}\;,\label{eq:NS}
\end{equation}
Here, $g$ is the standard gravity, and $\nu$ is the kinematic viscosity
of the fluid. The left-hand side of Eq.~(\ref{eq:NS}) is the advective
derivative of the vorticity. The two terms on the right-hand side
correspond to the diffusion of the vorticity due to internal friction
and to the buoyancy force, respectively.
{ Note that we have omitted in Eq.~(\ref{eq:NS}) the force exerted 
by the bacteria on the fluid, which has been taken into account in other works 
(see for example \cite{Lushi2012}). This is justified since we are interested 
only in large-scale flow. While the force exerted by the flagella on the
fluid during the run-and-tumble motion locally creates a strong agitation 
of the fluid, this does not lead to any macroscopic flow since the forces
are averaged over many bacteria. The dominant body force term that triggers 
bioconvection thus is the buoyancy force.} 

At the bottom and side boundaries,
the fluid velocity and the stream function are zero (no slip nor penetration).
On the free surface between fluid and air, the vertical component
of the velocity ($u_{y}$) and the tangential friction force $\left(\partial u_{x}/\partial y\right)$
are set to zero\footnote{ In the presence  of a biofilm the fluid flow is stopped
as can be seen in fig. \ref{fig:11h10-fluid} and this boundary condition is of
little importance }. This implies that the stream function is constant
at every boundary, and we set its boundary value to be zero.
Numerically, we have discretized Eq.~(\ref{eq:NS}) using standard
finite-difference formulas on a staggered grid, and solved it with
an explicit Euler scheme in time. The equation (\ref{eq:streamFunction})
was solved using a successive over-relaxation method (SOR) \cite{FuidDyn}.

\subsubsection{Oxygen concentration field}

As mentioned earlier, bacteria accumulation potentially plays an important
role in biofilm formation. Since this accumulation is mainly driven
by aerotaxis, we need a proper description of the oxygen distribution
in the fluid. The oxygen concentration field $\Oxygen(x,y,t)$ is
governed by four processes: diffusion, transport through the air-water
interface, consumption by the bacteria and advection by the flow.
Its evolution equation is 
\begin{equation}
\frac{\partial\Oxygen}{\partial t}+\Speedfluid\cdot\overrightarrow{\nabla}\Oxygen=D_{O2}\Delta\Oxygen-\gamma n\frac{\Oxygen}{\Oxygen+K}\;.\label{eq:oxygen}
\end{equation}
Here, $\Speedfluid\cdot\overrightarrow{\nabla}\Oxygen$ describes
the advection of the oxygen by the fluid, and $D_{O2}$ is the diffusion
coefficient of oxygen in water. The consumption of oxygen by the bacteria
at a concentration $n$ is modeled by a Michaelis-Menten law, which is one of the basic models
for enzymatic reactions. Oxygen is consumed at a constant rate $\gamma$
by each bacterium for concentrations much larger than the Michaelis
constant $K$. There are two boundary conditions for the oxygen,  no
flux at the walls (Neumann) and a constant concentration on the free
surface (Dirichlet): on the top surface, the oxygen concentration
is set to $C_{0}$ which is the saturation concentration of oxygen
in water for standard conditions. Equation (\ref{eq:oxygen}) is discretized 
on the same grid as the fluid flow equations, and integrated in time
using an explicit forward Euler scheme.

\subsection{Bacteria}

Each bacterium is represented as a discrete object and is characterized
by a number of variables: its instantaneous position and velocity,
the total mass (size) and internal state variables that indicate the
phenotype (motile or matrix producer), the connectivity (for the matrix
producer phenotype), and the time-integrated oxygen concentration
in the environment. In the following, we first describe the details
of bacterial motion: a random walk biased toward oxygen-rich regions
in space. Afterwards, we will describe the interactions between bacteria,
\textit{quorum sensing} and the handling of mechanical contacts. For the sake
of computational simplicity, the bacteria are considered to be spheres
(circles in two dimensions). Although \emph{Bacillus subtilis} has
a rod-like shape, this approximation should not lead to a qualitative
change in the behavior of the model in the initial stages of biofilm
formation.

\subsubsection{Random walk}

Bacteria are \emph{self-propelled} objects. The counter-clockwise
rotation of a flagella bundle creates a propulsion force that we consider
to be of constant magnitude. It drives the bacterium forward along
an almost straight trajectory with constant velocity of the order of 20 $\mu$m/s.
From time to time, the  { flagellum's} sense of rotation is inverted,
which leads to a rapid re-orientation of the { bacterium} body and 
to a change in the swimming direction. This
type of motion, which is common to many bacterial strains, is called
\emph{run-and-tumble} motion and can be well described as a random
walk.

During the run phase, with a typical Reynolds number of Re $=10^{-5}$, inertia can be neglected, and the bacterium move, relative to the fluid, with a constant velocity that results from a balance between the propulsion force $\ForcePropulsion$ and the viscous drag force $\Forcefriction$. Using Stokes' formula, for a spherical bacterium of velocity $\overrightarrow{\Speedbacteria}_{0}$ and radius $r$ the propulsion force is : 
\begin{equation}
\ForcePropulsion=-\Forcefriction=6\pi\eta r\overrightarrow{\Speedbacteria}_{0}=6\pi\eta rv_{0}\left(\begin{array}{c}
\cos\theta\\
\sin\theta
\end{array}\right)\;,
\end{equation}
where $\eta$ is the viscosity of the medium, $v_{0}$ is the modulus of the swimming velocity (assumed to be constant), and $\theta$ is the swimming direction.

The tumble process is simply modelled by the fact that there is, at
each time step, a probability for a change in the direction of $\theta$.
More precisely, we consider that tumbling is a Poissonian process
\cite{Berg}, with a mean run duration $\TimeRun$, and the probability
that a tumble takes place during the time step $\TimeStep$ is given
by 
\begin{equation}
\ProbabilityTumble=\frac{\TimeStep}{\TimeRun}\;.
\end{equation}
We assume that a tumble is instantaneous, and that the new direction
is randomly selected with a uniform probability distribution.

\subsubsection{Aerotaxis}

The run-and-tumble process decribed above generates an isotropic random
walk. In order to model aerotaxis, the tumble probability is modulated.
Motivated by the description of chemotaxis in \emph{Salmonella typhimurium}
\cite{Macnab1972} or in \emph{Escherichia coli} \cite{Segall1986,Keller1971},
we assume that the bacterium can keep track of the average oxygen
concentration over both a short ($\TimeMemoryShort$) and a long ($\TimeMemoryLong$)
time scale. This can be described by the use of two internal variables
$\MemoryShort$ and $\MemoryLong$ which obey the following equations:
\begin{eqnarray}
\frac{d\MemoryShort}{dt} & = & \frac{\Oxygen(\vec{x}(t),t)-\MemoryShort}{\TimeMemoryShort}\,,\label{eq:shortMeme}\\
\frac{d\MemoryLong}{dt} & = & \frac{\Oxygen(\vec{x}(t),t)-\MemoryLong}{\TimeMemoryLong}\,.\label{eq:longMeme}
\end{eqnarray}
Here, $\Oxygen(\vec{x}(t),t)$ is the oxygen concentration at time
$t$ at the position $\vec{x}(t)$ of the bacterium.  $\MemoryShort$ and
$\MemoryLong$ { can be seen respectively as an instantaneous measure of the local
oxygen  concentration and a fading memory of this quantity (similarly to the
work of \cite{Segall1986,Schnitzer1993}) or } as  the running averages over the oxygen concentrations encountered
by the bacterium over the time intervals $\TimeMemoryLong$ and $\TimeMemoryShort$,
respectively. 

If the bacterium swims in a favorable direction (toward oxygen), $\MemoryShort>\MemoryLong$ and $\ProbabilityTumble$ should decrease. Specifically, we have chosen
\begin{equation}
\ProbabilityTumble=\frac{\TimeStep}{\TimeRun}\frac{1}{1+\CoefAer(\MemoryShort(t)-\MemoryLong(t))}\,,\label{eq:proba aero}
\end{equation}
where $\CoefAer$ is a coefficient which sets the strength of the aerotaxis effect. As will be shown below, this simple model leads to a drift of the bacteria along the oxygen gradient. The parameter $\CoefAer$ is proportional to the coupling coefficient between the flux of bacteria and the oxygen gradient that is used in many continuum models to describe chemotaxis.

\subsubsection{Quorum Sensing}

The fact that biofilm formation takes place when the bacterial concentration
has reached a threshold indicates that the bacteria can, in some way,
sense their local concentration. It is believed that this mechanism,
called \textit{quorum sensing,} involves small molecules that are
both emitted and detected by the bacteria \cite{Bassler2006}. The
concentration of these molecules is thus a proxy for the bacteria
concentration in the vicinity. 

Here, in order to avoid the introduction of another concentration
field, we implement a probabilistic switch mechanism \cite{Chai2008,Lopez2010a}
using the density field $\BacterialCon$ introduced
in Sec.~\ref{sec_density}. A bacterium switches from the motile
to the matrix-producer phenotype with a probability 
\begin{equation}
\ProbabilityPhenotype=\left\{ \begin{array}{ll}
0 & \mbox{ if }n<n_{{\rm ph}}\\
\frac{\TimeStep}{\TimePheno} & \mbox{ if }n>n_{{\rm ph}},
\end{array}\right.\label{eq:proba threshold}
\end{equation}
where $\BacterialCon_{{\rm ph}}$ is the typical value of the bacterial concentration at which the switch from motile to matrix-producer phenotype occurs, and $\TimePheno$ is a characteristic time over which the phenotype change takes place.

\subsubsection{Motile bacteria: growth and contact}

As already mentioned, bacteria in their motile state exhibit a diffusive motion biased toward oxygen-rich regions. In addition, they grow and divide, which means that each bacterium (of index $i$) is also characterized by its radius $r_{i}$, from which one can compute its volume $V_{i}$ (we recall that we model bacterium as spherical objects): 
\begin{equation}
V_{i}=\frac{4}{3}\pi\Radius_{i}^{3}\;.\label{eq:volume}
\end{equation}
The volume of a bacterium evolves over time according to 
\begin{equation}
\frac{dV_{i}}{dt}=\frac{\VolumeDiv-V_{i}}{\TimeDiv}\frac{\Oxygen}{\Oxygen+\CoupureVol}\,,\label{eq:growthDiv}
\end{equation}
which corresponds to the growth toward $\VolumeDiv$ of the 
bacterial body size when it's ready to divide with a characteristic 
time $\TimeDiv$. When the oxygen concentration is small compared 
to $\CoupureVol$, growth virtually stops. Combining this growth 
law with Eq.~(\ref{eq:volume}), we obtain the evolution equation
of the bacterial radius: 
\begin{equation}
\frac{d\Radius_{i}}{dt}=\frac{1}{3\TimeDiv\,\Radius_{i}^{2}}(\RadiusDiv^{3}-\Radius_{i}^{3})\frac{\Oxygen}{\Oxygen+\CoupureVol}\,.\label{eq:eqDivRadius}
\end{equation}

Division is described by a random process: bacteria have a probability
of splitting into two that is a function of the local oxygen
concentration: 
\begin{equation}
\ProbabilityDivision=\frac{\TimeStep}{\TimeDiv}\frac{\Oxygen}{\Oxygen+\CoupureVol}\;.
\end{equation}
When a bacterium divides, it splits into two daughter cells. Imposing
volume conservation together with a spherical shape would imply that
each daughter cell has a radius of $1/\sqrt[3]{2}$ times the original
one. Therefore, the sum of the diameters of the daughters would be
larger than the diameter of the original bacterium, which would lead
to unphysical high repulsive forces between bacteria in a crowded
environment. Therefore, in the model, daughter cells have a diameter
that is equal to half the diameter of their mother, which avoids any
overlap induced by cell division. { Hence, during each cell 
division event there is a loss of bacterial volume. Nevertheless, one 
should keep in mind that the biomass is globally not conserved during 
growth (it increases with time). The loss of volume due to the method 
used for the division is necessary to ensure physical values of the 
contact forces between two neighboring bacteria under the constraint
of spherical shapes; this loss is compensated by the subsequent growth
of the two daughter bacteria so that, on average, biomass is increasing
as it should.}

Let us now describe the mechanical interactions between bacteria.
Since we neglect hydrodynamic interactions, for motile bacteria
there is only a soft-core repulsion ({bacteria that are in contact, i.e. when
$d_{ij}<r_{ij}$, repel each other}). We
model this interaction by a Lennard-Jones pair potential between particles
when they are close. This gives rise to the interaction force between
two bacteria $i$ and $j$ 
\begin{equation}
\ForceLennardJonnes^{ji}=\left\{ \begin{array}{ccc}
F_{0}\left(-\left(\frac{r_{ij}}{\DistanceEntreParticle_{ji}}\right)^{7}+\left(\frac{r_{ij}}{\DistanceEntreParticle_{ji}}\right)^{13}\right)\overrightarrow{\frac{\DistanceEntreParticle_{ji}}{\DistanceEntreParticle_{ji}}} & \textrm{if} & \frac{r_{ij}}{\DistanceEntreParticle_{ji}}>1\\
0 & \textrm{if} & \frac{r_{ij}}{\DistanceEntreParticle_{ji}}<1
\end{array}\right.
\end{equation}
where $\DistanceEntreParticle_{ij}$ is the distance between the centers
of the two neighboring bacteria, $\Radius_{ij}=\Radius_{i}+\Radius_{j}$
is the sum of their radii, and $F_{0}$ is a parameter that describes
the intensity of the force.

\subsubsection{Matrix producers: growth and contact}

When the bacteria change to the matrix-producer phenotype, they stop
dividing, stop propelling, and start to produce extracellular matrix
which allows them to bind to other bacteria (motile or matrix producers).
Ultimately, the bacteria are embedded in a soft elastic medium. This
makes the mechanical interactions between bacteria complicated and
non-local. Here, we make drastic simplifying assumptions to make the
model tractable, while the essential features of the biofilm material
are reproduced.

The matrix producers do not divide, and their propulsion force $\ForcePropulsion$
is set to zero. Matrix production is modeled by an increase of the
particle volume over time: 
\begin{equation}
\frac{dV_{i}}{dt}=\frac{\VolumeMax-V_{i}}{\TimeMatrix}\frac{\Oxygen}{\Oxygen+\CoupureVol}\,,
\end{equation}

which corresponds to a finite amount $\VolumeMax$ of matrix produced by each bacterium within the characteristic time $\TimeMatrix$. As a result, the radius of the matrix producer grows according to 
\begin{equation}
\frac{d\Radius_{i}}{dt}=\frac{1}{3\TimeMatrix\,\Radius_{i}^{2}}(r_{m}^{3}-\Radius_{i}^{3})\frac{\Oxygen}{\Oxygen+\CoupureVol}\,.\label{eq:eq augmentation volum}
\end{equation}
In addition to the repulsive force previously described, we consider that once a bacterium (matrix producer or motile) is in contact with a matrix producer, a link between them is established. More precisely, as soon as the distance between the centers of the two bacteria $d_{ij}$ becomes smaller than the sum of their radii $\Radius_{ij}$, this \foreignlanguage{english}{\textit{link} is established. We model the links as simple linear springs with spring constant $k$ that break when the distance between the bacteria is larger than $2\Radius_{ij}$. When these forces are combined with the hard-core repulsion, the inter-bacterial interaction writes finally:
\begin{equation}
\ForceLennardJonnes^{ji}=\left\{ \begin{array}{ccc}
F_{0}\left(-\left(\frac{r_{ij}}{\DistanceEntreParticle_{ji}}\right)^{7}+\left(\frac{r_{ij}}{\DistanceEntreParticle_{ji}}\right)^{13}\right)\overrightarrow{\frac{\DistanceEntreParticle_{ji}}{\DistanceEntreParticle_{ji}}} & \textrm{if} & \DistanceEntreParticle_{ji} \leq \Radius_{ij}\\
-k\,(\DistanceEntreParticle_{ji}-\Radius_{ij})\overrightarrow{\frac{\DistanceEntreParticle_{ji}}{\DistanceEntreParticle_{ji}}} & \textrm{if} & \Radius_{ij}<\DistanceEntreParticle_{ji} \leq 2\,\Radius_{ij}\\
0 & \textrm{if} & 2\Radius_{ij}<\DistanceEntreParticle_{ji}
\end{array}\right.\label{eq:force_elastique}
\end{equation}
With these simple rules, we can capture the steric exclusion generated by the finite volume of the bacterium, as well as elastic and plastic deformation of the biofilm.}

\subsubsection{Velocity calculation and boundary conditions}

Since the flow of the fluid medium around a bacterium is characterized by a Reynolds number much smaller than unity,
bacterial motion is the result of the balance between propulsion (for motile bacteria), viscous drag, and interaction forces. 
This writes: 
\begin{equation}
\sum_{j\,{\rm contacts}}\ForceLennardJonnes^{ji}+\overrightarrow{f}_{p}^{i}-6\pi\eta\Radius_{0}\overrightarrow{\Speedbacteria}_{i}=\overrightarrow{0}\;.\label{eq:bacterial velocity}
\end{equation}
where the propulsion force is set to zero for matrix producers. Here,
$\overrightarrow{\Speedbacteria}_{i}$ is the velocity of the bacterium $i$ with
respect to the fluid. In order to obtain its velocity in the laboratory frame,
the local fluid flow velocity must be added. { The rotation of 
the bacteria induced by fluid flow is neglected since it takes place on 
a time scale that is much larger than a typical bacterial run length
(in the experiments, bioconvection typically took place on the millimeter
scale with a velocity of the order of 1 $\mu$m/s, which yields a shear
rate of $\sim 10^{-3}$ s$^{-1}$).}

Boundary conditions for the bacteria also need to be specified. On
the side walls, to ensure adhesion of the biofilm, immobile planktonic
bacteria are disposed so that bacteria in the matrix producer state
are able to bind to the wall. In addition, the repulsive Lennard-Jones
force prevents motile bacteria from crossing the walls. On the air-water
interface that is supposed to remain flat, motile bacteria are assumed
to reverse their propulsion so that they cannot cross the surface.
However, bacteria can be ``pushed'' beyond the surface under the
action of contact forces.

\section{Choice of parameters}

\label{sec_params} The model presented above contains a number of
parameters. The ones related to physical processes, such as the viscosity
of the medium, are known with good precision. In contrast, parameters
of biological processes (such as the oxygen consumption of a bacterium)
are often known only with large error bars. Finally, some model parameters,
such as the strength of the interactions between bacteria, appear
in approximations that are specific to this model. Therefore, they
cannot be measured directly, but must be estimated from macroscopic
properties. The motivations for most of our choices are discussed in the appendix,
and the values of the parameters are summarized in table \ref{tab:parameters}.
Here, we discuss our choices for the aerotaxis parameter $\CoefAer$
and the spring constant $k$, since they require some further analysis.

\begin{sidewaystable}
{\footnotesize
\begin{tabular}{|>{\centering}p{7cm}|c|c|>{\centering}p{1cm}|c|}
\hline 
Name  & Symbol  &  & Eq  & Ref\tabularnewline
\hline 
\hline 
Rate of oxygen consumption  & $\gamma$  & \textbf{$2\cdot10^{6}$ }$\textrm{\,\ molecules}\cdot\textrm{s}^{-1}\cdot\textrm{bacterium}{}^{-1}$  & \ref{eq:oxygen}  & \cite{Martin1932,Douarche2009_PRL}\tabularnewline
\hline 
Michaelis constant  & $K$  & $10^{-3}\,\textrm{molecules}\cdot\textrm{m}^{-3}$  & \ref{eq:oxygen}  & \tabularnewline
\hline 
Spring constant  & $k$  & $10^{-8}\,\textrm{N}\cdot\textrm{m}^{-1}$  & \ref{eq:force_elastique}  & \tabularnewline
\hline 
Lennard Jones coefficient  & $F_{0}$  & $10^{-5}$ N & \ref{eq:force_elastique}  & \tabularnewline
\hline 
Bacterial speed  & $\Speedbacteria_{0}$  & $20\,\textrm{\ensuremath{\mu}m}\cdot\textrm{s}^{-1}$  & \ref{eq:bacterial velocity}  & \cite{Cisneros2008these}\tabularnewline
\hline 
Threshold for the phenotype switch  & $n_{ph}$  & $2.10^{14}\,\textrm{bacteria}\cdot\textrm{m}^{-3}$  & \ref{eq:proba threshold}  & \tabularnewline
\hline 
Rate of the phenotype switch  & $\TimePheno$  & $\Delta t$  & \ref{eq:proba threshold}  & \tabularnewline
\hline 
Mass density of the bacteria  & $\rho_{b}$  & $1,03\,\textrm{kg}\cdot\textrm{m}^{-3}$  & \ref{eq:NS}  & \tabularnewline
\hline 
Radius of the bacteria  & $\Radius_{0}$  & $5\,\textrm{\ensuremath{\mu}m}$  & \ref{eq:NS}  & \tabularnewline
\hline 
Mean duration of a run in homogenous environment  & $\TimeRun$  & $1$ s  & \ref{eq:proba aero}  & \cite{Berg} \tabularnewline
\hline 
Coefficient of aerotaxis  & $\CoefAer$  & $10^{-22}\textrm{\textrm{m}}^{3}\textrm{\ensuremath{\cdot}molecule}{}^{-1}$  & \ref{eq:proba aero}  & \tabularnewline
\hline 
Maximal radius of the  bacteria  & $\RadiusDiv$  & $\sqrt[3]{2}\Radius$  & \ref{eq:growthDiv}  & \tabularnewline
\hline 
Rate of a bacterial body increase  & $\TimeDiv$  & $70$ min  & \ref{eq:growthDiv}  & \tabularnewline
\hline 
Michaelis saturation of the growth rate & $\CoupureVol$ & $10^{-2}\,\textrm{molecules}\cdot\textrm{m}^{-3}$ & \ref{eq:growthDiv} & \tabularnewline
\hline 
Maximal radius of the matrix producer  & $\Radius_{m}$  & $5\Radius_0$  & \ref{eq:eq augmentation volum}  & \tabularnewline
\hline 
Rate of matrix production  & $\TimeMatrix$  & $1$ h  & \ref{eq:eq augmentation volum}  & \tabularnewline
\hline 
Integration time of the oxygen short memory  & $\TimeMemoryShort$  & $0,1$ s  & \ref{eq:shortMeme}  & \cite{Wong1995}\tabularnewline
\hline 
Integration time of the oxygen long memory  & $\TimeMemoryLong$  & $10$ s  & \ref{eq:longMeme}  & \cite{Wong1995}\tabularnewline
\hline 
Fluide viscosity & $\eta$ & $10^{-3}\,\textrm{Pa}^{2}\cdot\textrm{s}^{1}$ &  & \tabularnewline
\hline 
Grid size for concentration  & $\Delta x$  & $5.10^{-4}$ m  &  & \tabularnewline
\hline 
Thickness of the 2D slice  & $dz$  & $2\Radius_0$  &  & \tabularnewline
\hline 
Width and height of the 2D slice  & $Lx$  & $10^{-2}$ m  &  & \tabularnewline
\hline 
\color{red}{Oxygen boundary condition}  & $C_{0}$  & $1.5\, 10^{23} \, \textrm{molecules}/\textrm{m}^{3}$  &   & \tabularnewline

\hline 
\end{tabular}
}
\protect\caption{Summary of the parameters\label{tab:parameters} }
\end{sidewaystable}

\subsection{Aerotaxis}

Our goal is to relate the parameter $\CoefAer$ to the more conventional
descriptions of taxis by partial differential equations. This is a
classic subject (see for example Ref.~\cite{Alt1980} for a detailed
exposition), and we give here only a few elements of the analysis
as applied to our specific model. Consider an ensemble of non-interacting
bacteria that perform a run-and-tumble motion in a uniform gradient
of oxygen concentration $G_{0}$ along the $y$ direction, that is,
\begin{equation}
c(\vec{x})=c_{0}+G_{0}y,\label{eq:oxygenG}
\end{equation}
where $c_{0}$ is a reference concentration at the position $y=0$.
For a bacterium that runs along a straight line (without tumbles),
Eqs.~(\ref{eq:shortMeme}) and (\ref{eq:longMeme}) for the internal
memory variables can be solved exactly, 
\begin{equation}
m_{s,l}(t)=v_{0}\cos\theta G_{0}t+\left(c_{0}-v_{0}\cos\theta G_{0}\tau_{s,l}\right)+A_{s,l}\exp(-t/\tau_{s,l})\label{eq:exact}
\end{equation}
where the constants $A_{s,l}$ are determined by the initial conditions
for $m_{s,l}$ and are unimportant in the long-time limit, and $\theta$
is the angle between the run direction and the $y$ axis. This yields
(for long times) 
\begin{equation}
m_{l}-m_{s}=v_{0}\cos\theta G_{0}(\tau_{l}-\tau_{s})\label{eq:deltamav}
\end{equation}
For a real trajectory of a bacterium (with tumbles), the time evolution
of the memory variables is more complicated. However, since $\tau_{s}$
is much shorter than the average duration of a run, $m_{s}$ will
be close to the solution of Eq.~(\ref{eq:exact}). In contrast, since
$\tau_{l}$ is much longer than the run duration, the slow variable
will average over several run directions. In a whole population of
bacteria, the mean probability for tumbling therefore depends only
on the current run direction $\theta$ (through the fast variables).
We denote this probability density by $\mathcal{P}(\theta)$. Furthermore,
the time average of $m_{l}-m_{s}$ is proportional to the right-hand
side of Eq.~(\ref{eq:deltamav}). Since the tumble probability must
decrease for a favorable run direction, we have to first order in
$G_{0}$ 
\begin{equation}
\mathcal{P}(\theta)\approx\frac{1}{\TimeRun}\left(1-\alpha\TimeAjustable\OxygenGradient v_{0}\cos\theta\right)\approx\frac{1}{\TimeRun}\frac{1}{1+\alpha\TimeAjustable\OxygenGradient v_{0}\cos\theta}\,,\label{eq:probaTumb_continu}
\end{equation}
with $\TimeAjustable$ a parameter of dimension time.

The knowledge of the probability for a bacterium to swim in the direction
$\theta$ at time $t$ $\ProbabilityDensityAngle(\theta,t)$ is sufficient
to determine the global motion of the bacteria population. Since there
is no persistence during a tumble, $\ProbabilityDensityAngle(\theta,t)$
satisfies the simple master equation 
\begin{equation}
\partial_{t}\ProbabilityDensityAngle(\theta,t)=-\ProbabilityDensityAngle(\theta,t)\mathcal{P}(\theta)+\frac{1}{2\pi}\int_{-\pi}^{+\pi}\ProbabilityDensityAngle(\theta',t)\mathcal{P}(\theta')d\theta'\,,
\end{equation}
When the population is in steady state, the time derivative is zero, which implies that $\ProbabilityDensityAngle(\theta,t)\mathcal{P}(\theta)$ is constant. Since $\ProbabilityDensityAngle(\theta,t)$ is normalized by: 
\begin{equation}
\int_{-\pi}^{+\pi}\ProbabilityDensityAngle(\theta,t)d\theta=1,
\end{equation}
we obtain 
\begin{equation}
\ProbabilityDensityAngle(\theta)=\left(\int_{-\pi}^{+\pi}\frac{1}{\mathcal{P}(\theta')}d\theta'\right)^{-1}\frac{1}{\mathcal{P}(\theta)}\,,
\end{equation}
where time-dependence has been dropped for the ease of notation. With Eq.~(\ref{eq:probaTumb_continu}) we obtain 
\begin{equation}
\ProbabilityDensityAngle(\theta)=\frac{1}{2\pi}(1+\alpha\TimeAjustable\OxygenGradient v_{0}\cos\theta)\,.
\end{equation}
The average velocity (\textit{i.e.} the drift velocity) in the direction of the gradient can be written in terms of $\ProbabilityDensityAngle(\theta,t)$ as 
\begin{equation}
v_{drift}=\frac{1}{2\pi}\frac{\int_{-\pi}^{+\pi}v_{0}cos(\theta')\ProbabilityDensityAngle(\theta')d\theta'}{\int_{-\pi}^{+\pi}\ProbabilityDensityAngle(\theta')d\theta'}=\int_{-\pi}^{+\pi}v_{0}cos(\theta')\ProbabilityDensityAngle(\theta')d\theta'\,.\label{eq:vdrift_calcul}
\end{equation}
This yields finally 
\begin{equation}
v_{drift}=\frac{1}{2}\alpha\OxygenGradient\TimeAjustable v_{0}^{2}\,.\label{eq:vdrift}
\end{equation}

\begin{figure}[tbph]
\noindent \begin{centering}
\includegraphics[clip,width=13cm]{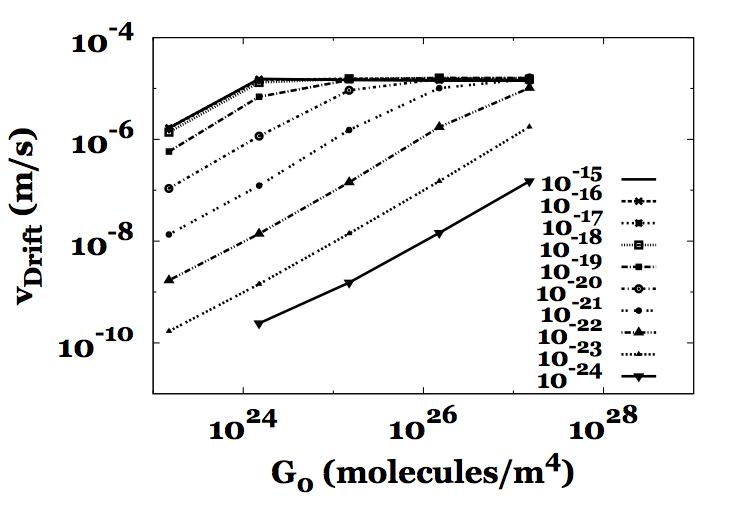} 
\par\end{centering}

\protect\caption{Drift velocity $\protect\Speedbacteria_{drift}$ calculated by simulation as a function of the oxygen gradient $\protect\OxygenGradient$ for different values of the aerotaxis coefficient $\protect\CoefAer$.
\label{fig:Drift-velocity}}
\end{figure}

\begin{figure}[H]
\noindent \begin{centering}
\includegraphics[clip,width=13cm]{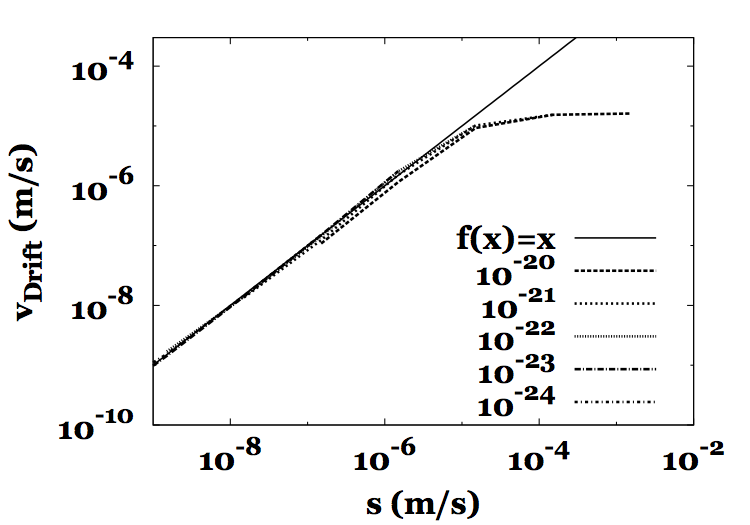} 
\par\end{centering}

\protect\caption{Drift velocity versus $\protect\CoefRenorm=\protect\OxygenGradient\protect\CoefAer\protect\Speedbacteria_{0}^{2}\protect\TimeAjustable/2$. The simulations for different values of $\alpha$ and $\protect\OxygenGradient$ collapse together, and the saturation of the drift velocity is near $G_{0}\alpha v_{0}=1$. $\protect\TimeAjustable=0.5\ s$.
\label{fig:collapseDrift}}
\end{figure}

To validate this prediction, numerical simulations were performed
with an oxygen concentration given by Eq.~(\ref{eq:oxygenG}) and
an ensemble of 10000 bacteria in an infinite medium. Different values
of $\OxygenGradient$ and $\CoefAer$ were considered. In figure \ref{fig:Drift-velocity},
the drift velocity, obtained by averaging over all the bacteria and
over long runs, is plotted against $\OxygenGradient$ for various
values of $\CoefAer$. One can see a linear increase until the drift
velocity reaches a plateau at the value of $v_{{\rm drift}}\approx\Speedbacteria_{0}=2.10^{-5}$
m/s. The plateau can be attributed to the finite swimming speed of
the bacteria: the drift velocity cannot exceed the swimming velocity.
As predicted by Eq.~(\ref{eq:vdrift}), after renormalization of
$\OxygenGradient$ by $\CoefAer$, the curves collapse on a master
curve as can be seen in Fig.~\ref{fig:collapseDrift}. The value
of $\TimeAjustable$ that is given by a fit corresponds to half
of the unbiased run time $\TimeRun$.

The model should exhibit a sizeable aerotaxis effect, but should not
be affected by spurious effects induced by the plateau in the curve
(the drift velocity should remain smaller than the swimming velocity).
Since the order of magnitude of the oxygen gradient found in our simulations
is at most $10^{26}$ molecules/m$^{4}$, this requires to choose
$\CoefAer<10^{-20}$ m$^{3}$/molecule.

\subsection{Spring constant}

An estimation for the ``spring constant'' $k$ of a
matrix ``bridge'' between bacteria can be deduced from the elastic
modulus of the biofilm, that has been measured recently \cite{Trejo2013}.
Assuming that each matrix link between two bacteria has a cross-section
of $\pi\Radius_{0}^{2}$ and an equilibrium length of $2\Radius_{0}$,
the stress is $\sigma=k\Delta r/(\pi r_{0}^{2})$, where $\Delta r$
is the elongation of the link. Using Hooke's law for an isotropic
medium of Young modulus $Y$, one finds that $\sigma=Y\Delta r/(2r_{0})$.
These two relations yield 
\begin{equation}
k\sim Y\frac{\pi}{2}r_{0}.
\end{equation}

If the measured value of the elastic modulus is used to calculate
the spring constant, a problem arises for the numerical simulations,
which is due to the multiple time scales present in the problem of
biofilm formation. Indeed, considering Eq.~(\ref{eq:bacterial velocity})
applied on two matrix-producer bacteria that are linked by a strained
matrix bridge in a fluid at rest, the bridge will relax to its equilibrium
length with a characteristic time scale $\tau$ given by:
\begin{equation}
\tau\sim\frac{6\pi\eta r_{0}}{k}=\frac{12\eta}{Y}.
\end{equation}
For the Young's modulus measured in biofilms ($Y=10-10000\, Pa$ 
\cite{Trejo2013}), this time scale ranges from $\tau=10^{-3}-10^{-6}$ s. 
The numerical integration requires a timestep that is much smaller
than this time scale in order to properly resolve the dynamics. However,
biofilm formation takes several days. Thus, it is extremely difficult
to perform simulations of biofilm formation for the physical values
of Young's modulus. Therefore, we have chosen to use much lower values
for the spring constant, corresponding to $Y<10^{-1}\, Pa$, which
implies that our simulated biofilms are less stiff than in reality.
Finally, for the repulsion, the parameter $F_{0}$ is taken to be
$10^{-5}\, N$ following similar considerations.

\section{Results}

\label{sec_results} We have performed numerical simulations
to test the behavior of our model. They show that the
model can reproduce all the main stages of biofilm formation.
In addition, systematic parametric studies have allowed us to identify
the model parameters that have the strongest influence on the biofilm
growth dynamics and morphology. Those are the division time of the
bacteria $\TimeDiv$ (for the timing of the biofilm nucleation), the
value of the bacterial concentration threshold for the phenotypic
switch $n_{ph}$, and the value of the spring constant $k$. Moreover,
we have tested the influence of bioconvection by comparing simulations
with and without coupling to fluid flow. 

In the following, we first present a \textit{reference} simulation 
in order to provide a description of the typical time evolution of 
the system. Then, the influence of several model parameters is 
discussed. All other parameters remain fixed to the values given 
in Table \ref{tab:parameters}.

\subsection{Steps of biofilm formation}

\begin{figure}[H]
\noindent 
\subfloat[\label{fig:7h50}]{\includegraphics[bb=112bp 15bp 524bp 430bp,clip,width=4cm]{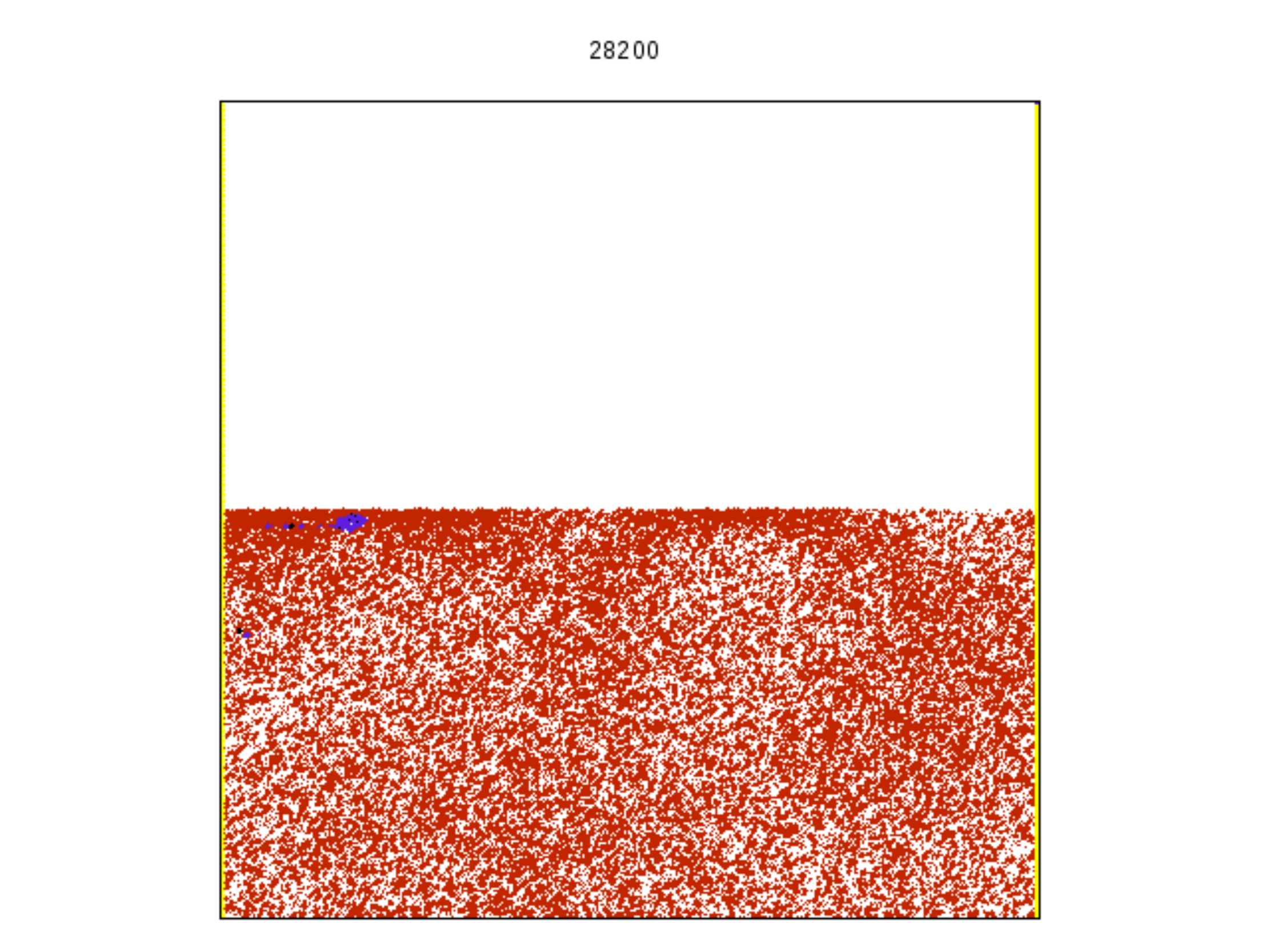}}
\subfloat[\label{fig:8h10}]{\includegraphics[bb=112bp 15bp 524bp 430bp,clip,width=4cm]{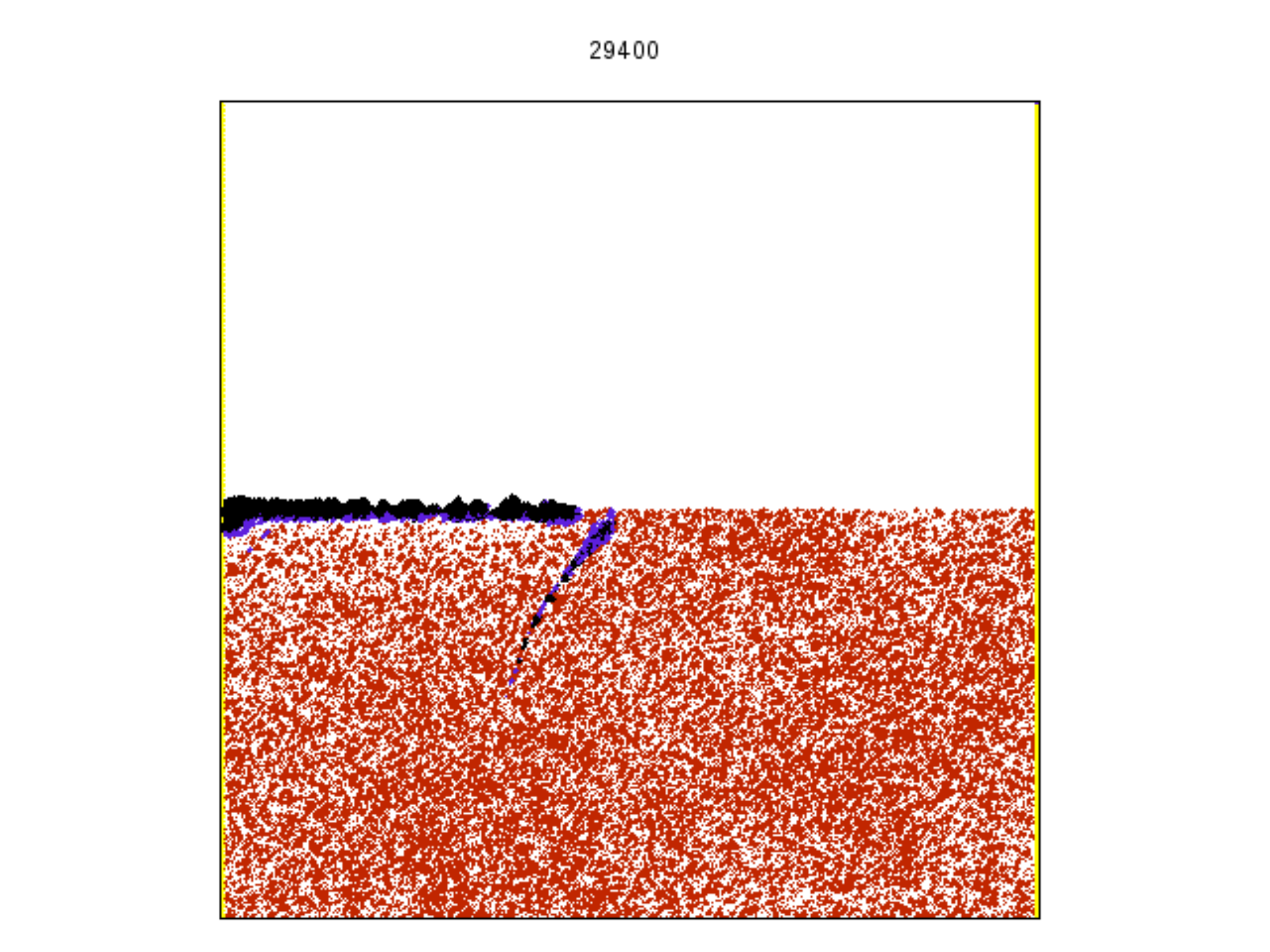}}
\subfloat[\label{fig:8h20}]{\includegraphics[bb=112bp 15bp 524bp 430bp,clip,width=4cm]{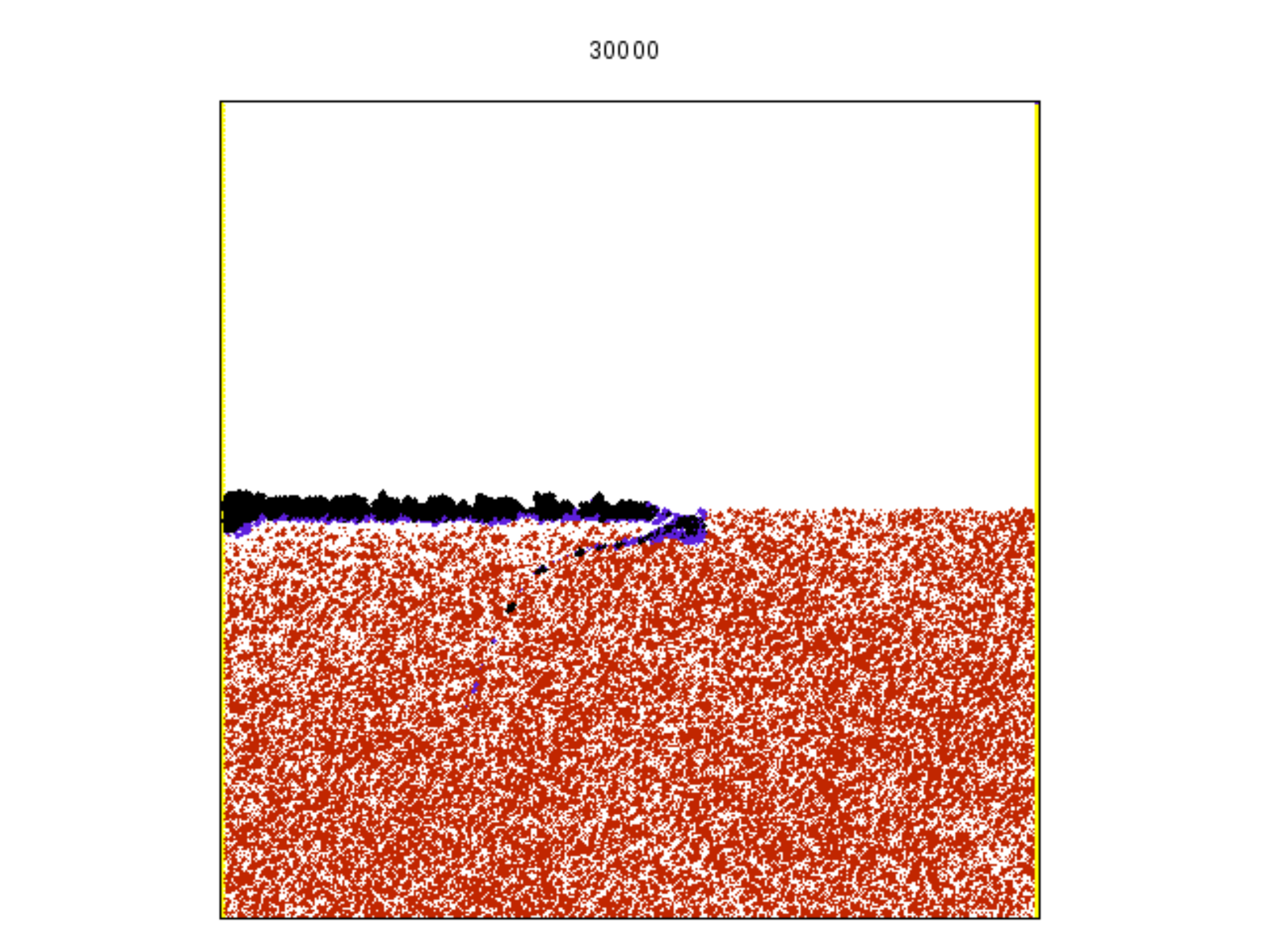}}
\subfloat[\label{fig:11h10}]{\includegraphics[bb=112bp 15bp 524bp 430bp,clip,width=4cm]{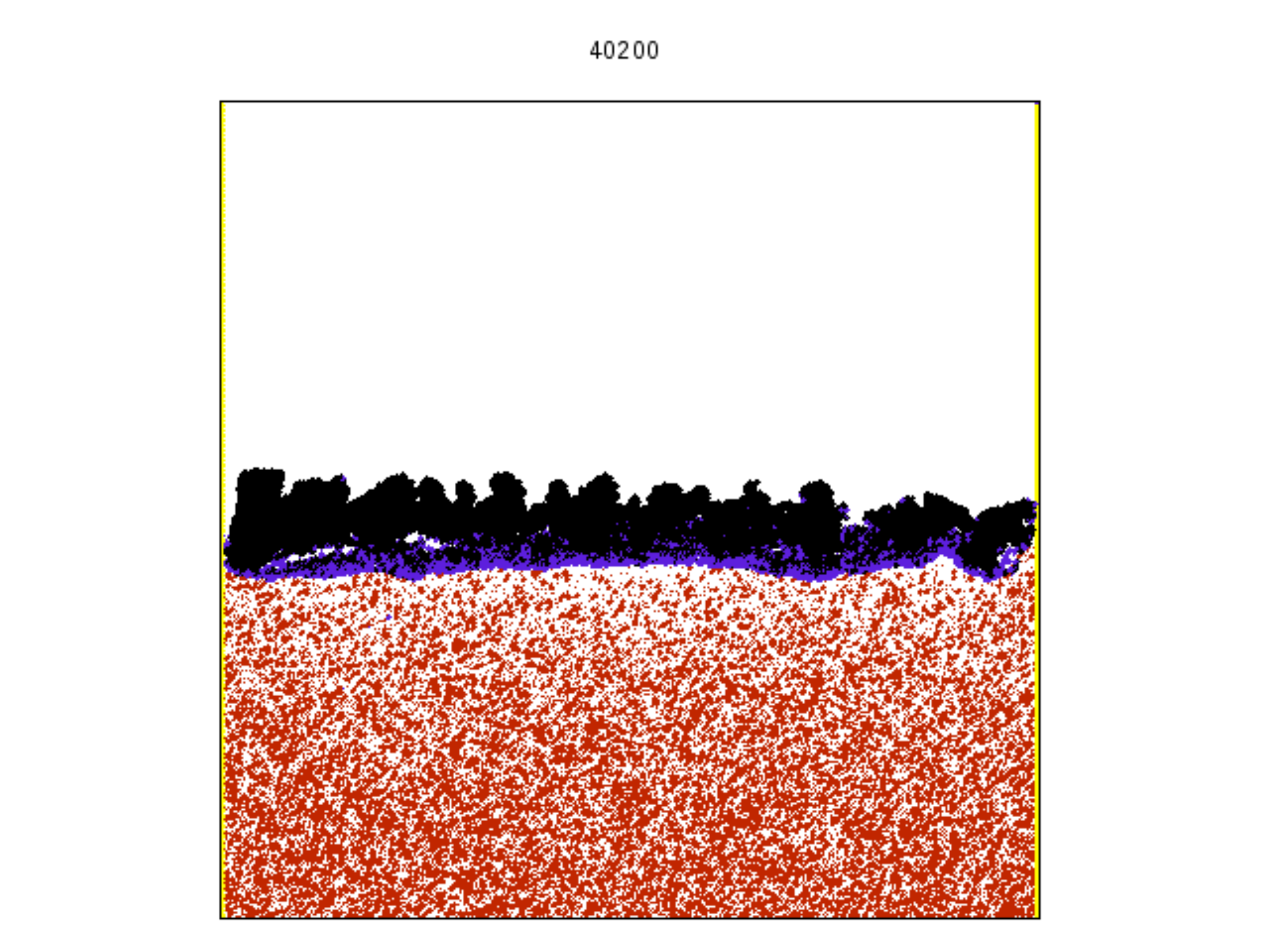}}

\noindent 
\subfloat[\label{fig:7h50-fluid}]{\includegraphics[bb=145bp 30bp 550bp 430bp,clip,width=4cm]{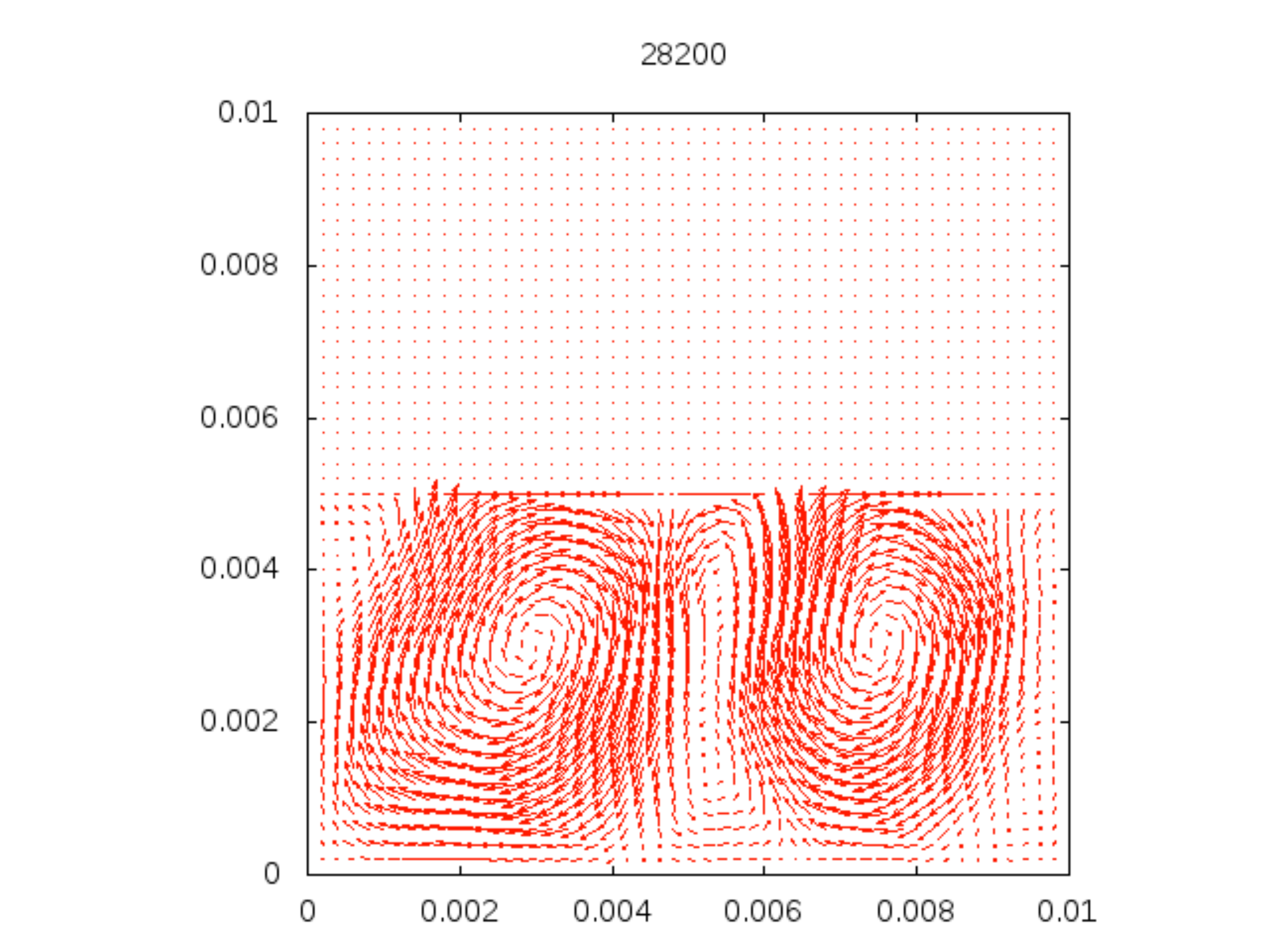}}
\subfloat[\label{fig:8h10-fluid}]{\includegraphics[bb=145bp 30bp 550bp 430bp,clip,width=4cm]{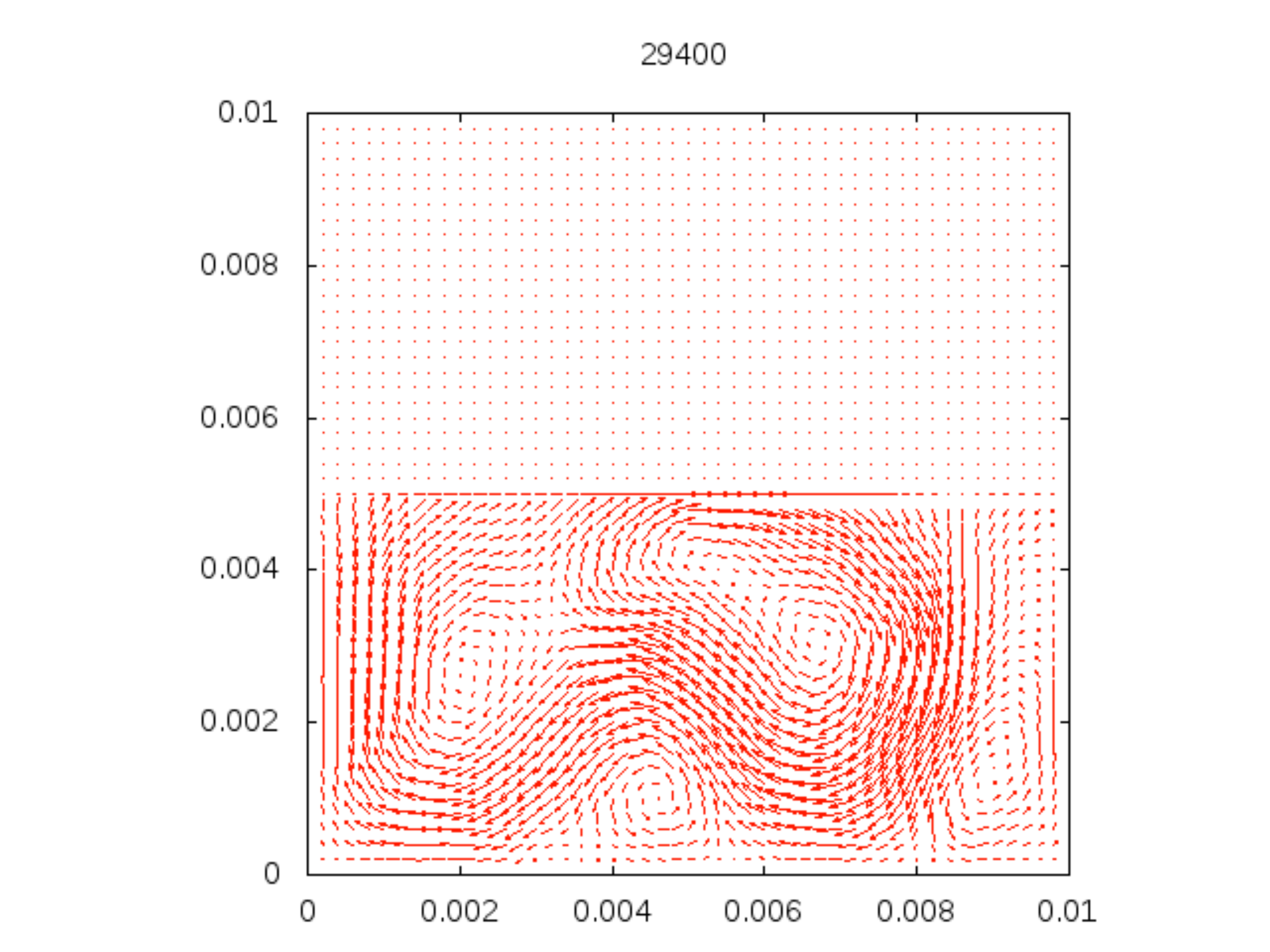}}
\subfloat[\label{fig:8h20-fluid}]{\includegraphics[bb=145bp 30bp 550bp 430bp,clip,width=4cm]{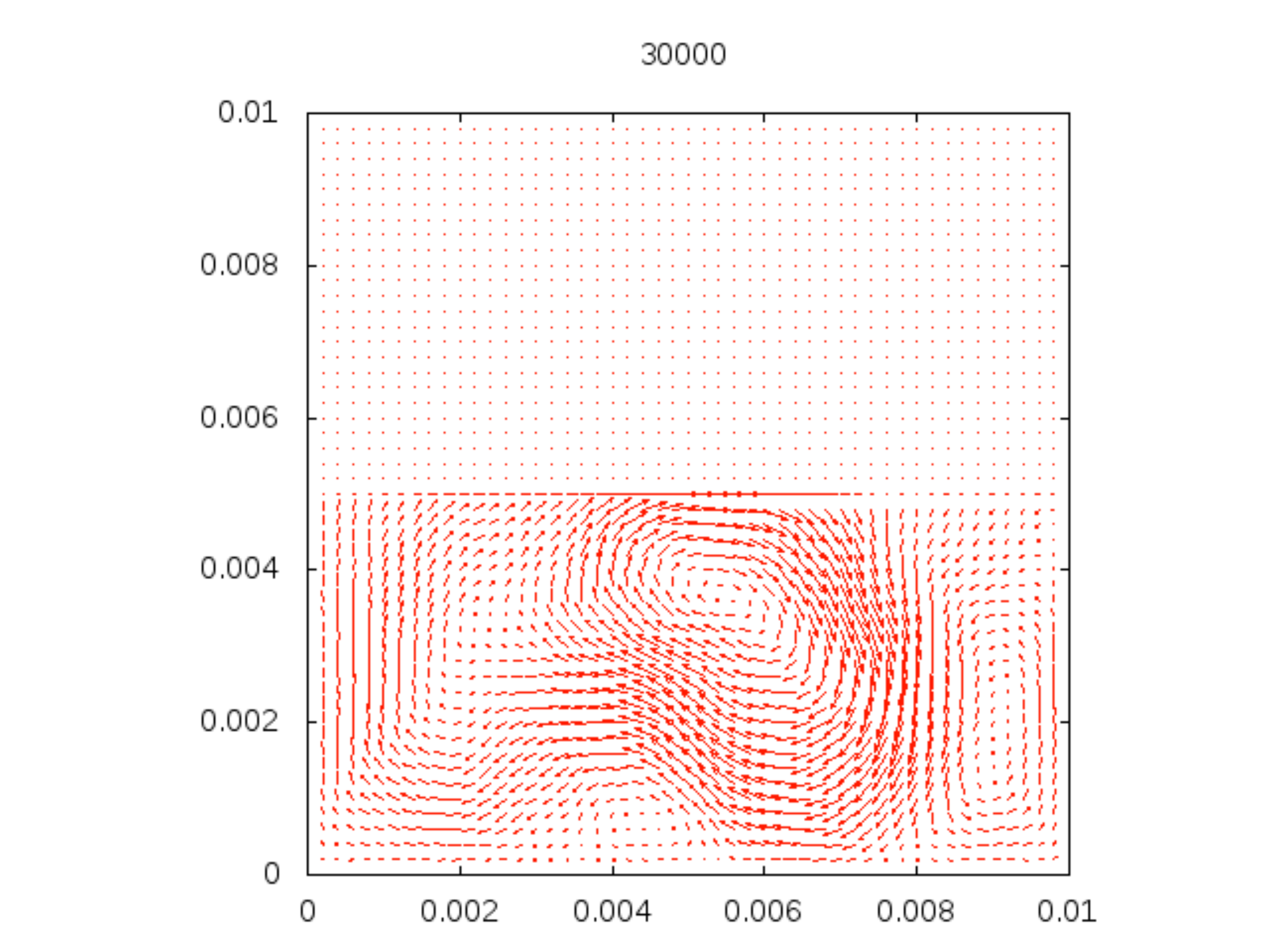}}
\subfloat[\label{fig:11h10-fluid}]{\includegraphics[bb=145bp 30bp 550bp 430bp,clip,width=4cm]{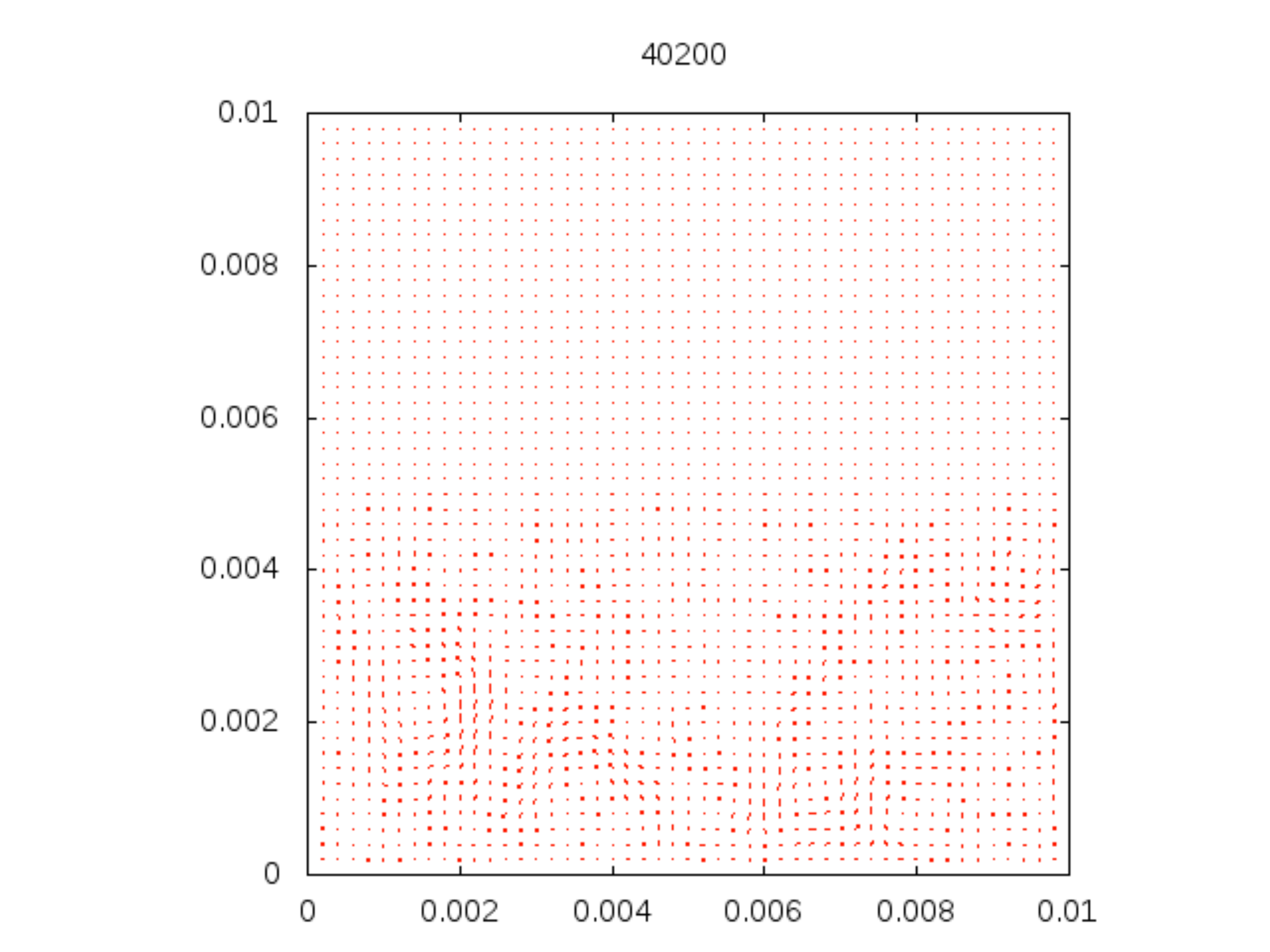}}

\noindent 
\subfloat[\label{fig:7h50-O2}]{\includegraphics[bb=165bp 97bp 470bp 400bp,clip,width=4cm]{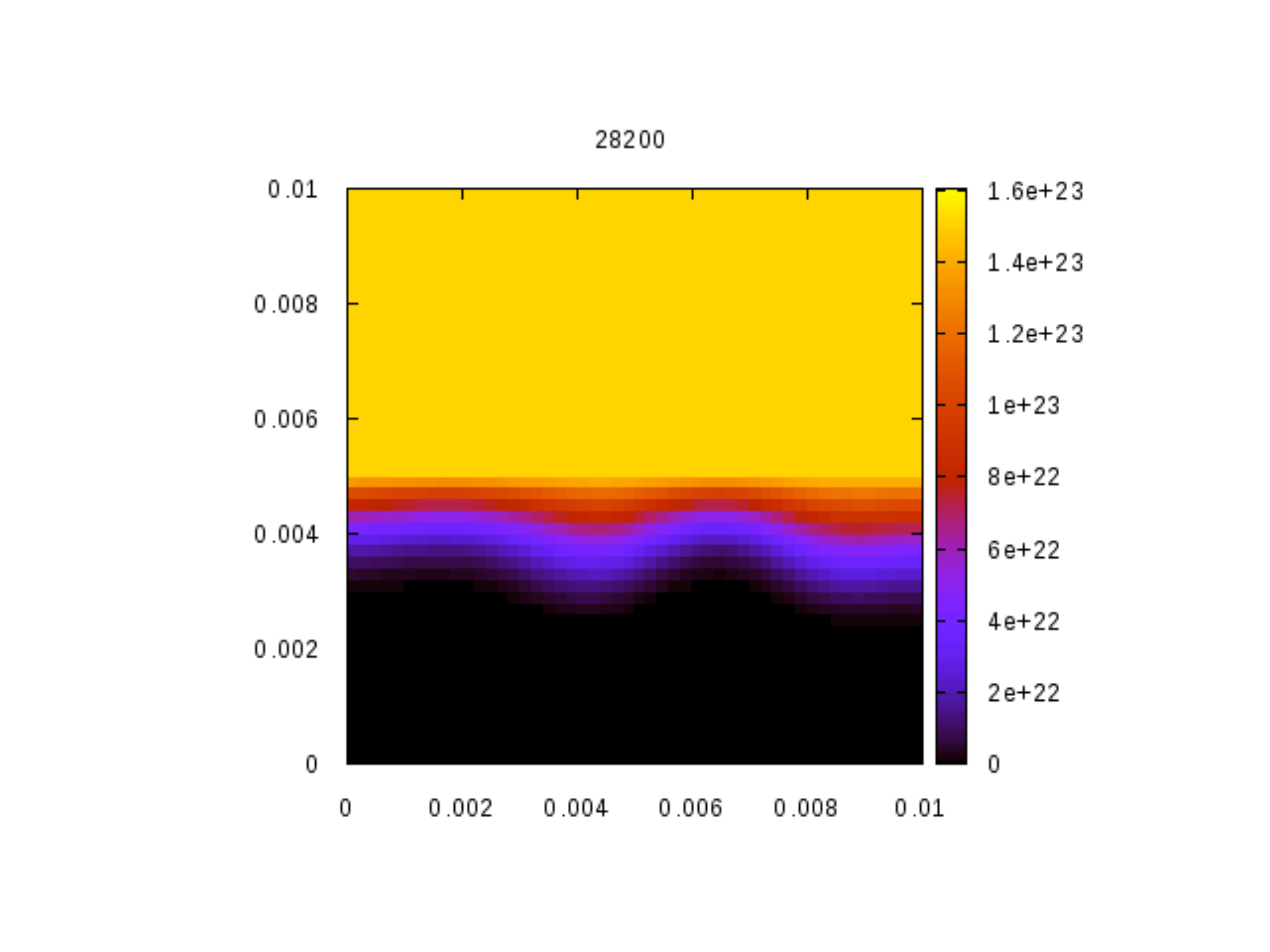}}
\subfloat[\label{fig:8h10-O2}]{\includegraphics[bb=165bp 97bp 470bp 400bp,clip,width=4cm]{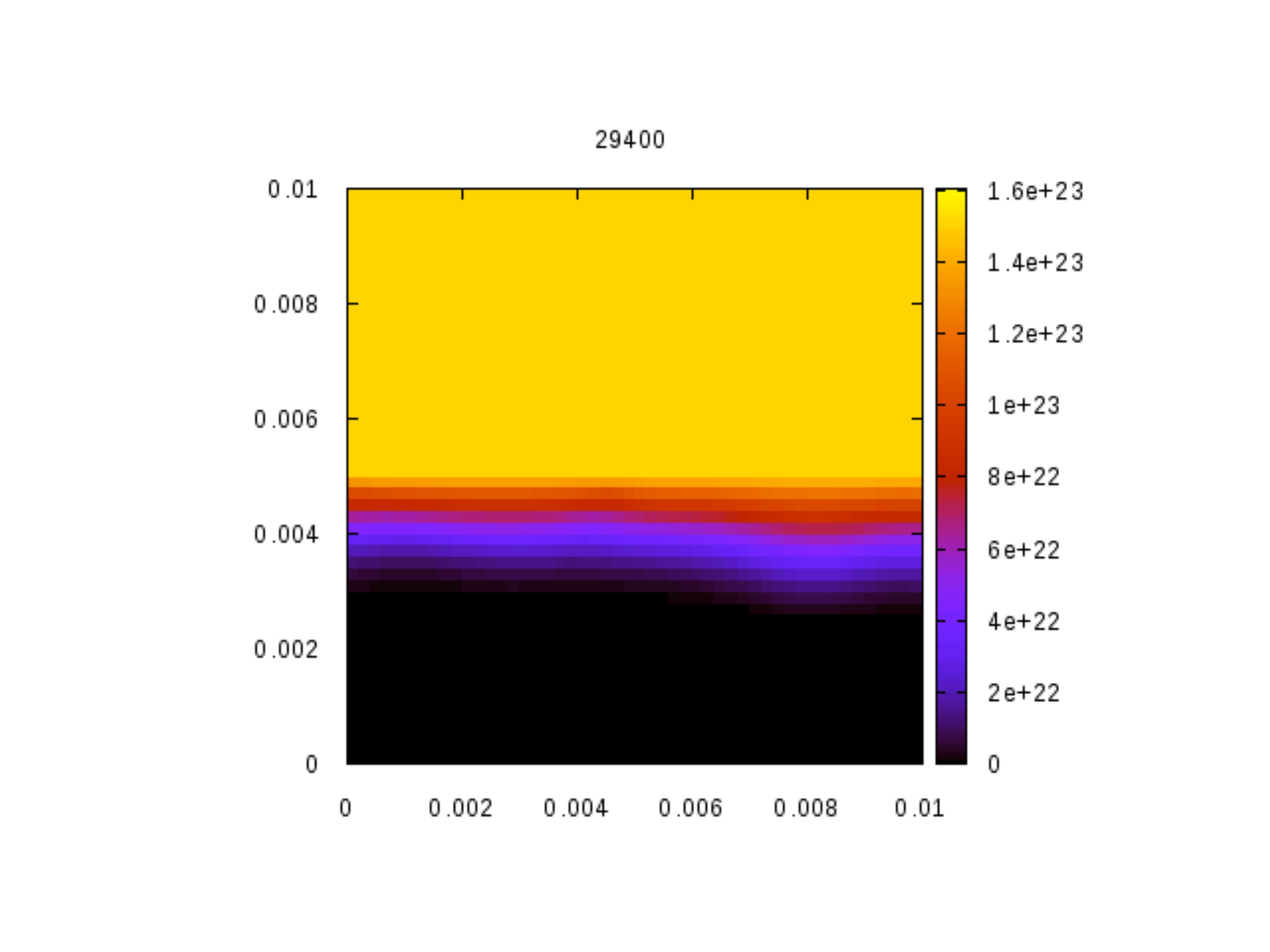}}
\subfloat[\label{fig:8h20-O2}]{\includegraphics[bb=165bp 97bp 470bp 400bp,clip,width=4cm]{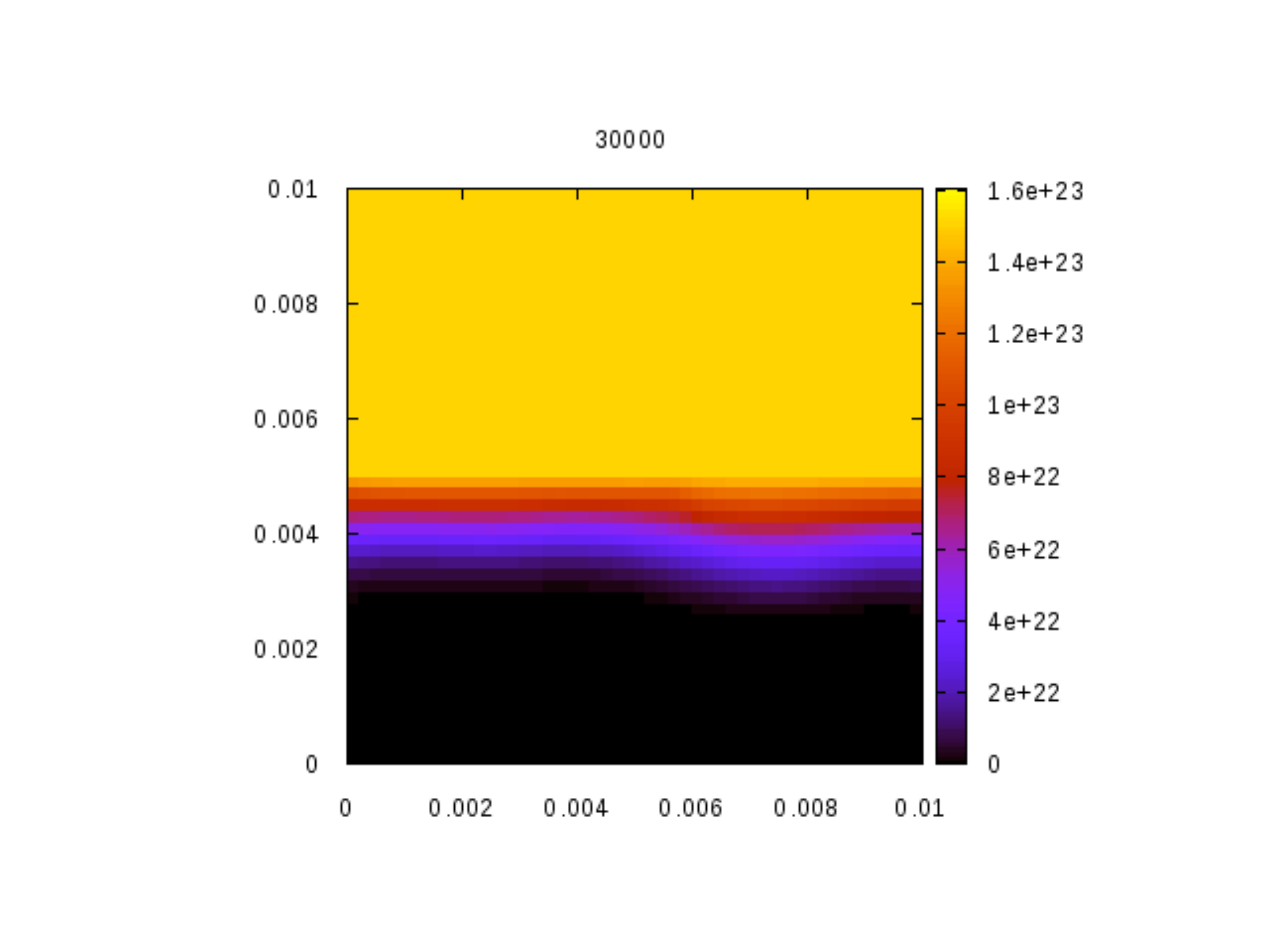}}
\subfloat[\label{fig:11h10-O2}]{\includegraphics[bb=165bp 97bp 490bp 400bp,clip,width=4.25cm]{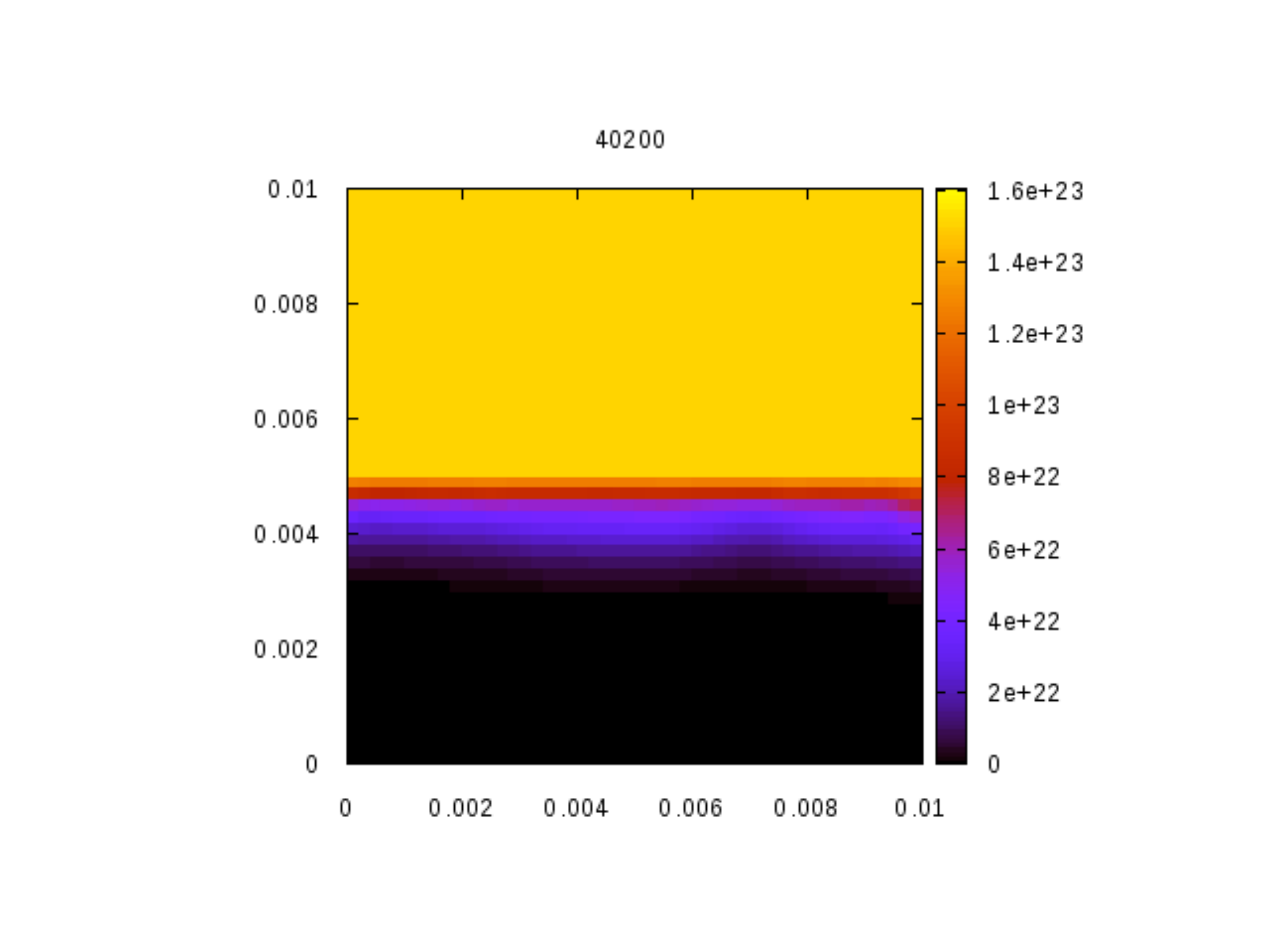}}

\protect\caption{The first line shows successive snapshots of the bacterial population
(at 7h50 (a), 8h10 (b), 8h20 (c) and 11h10 (d)) during the formation
of a biofilm. Second and third lines show respectively the evolution
of the fluid velocity and the oxygen. The parameters for the simulation
are the one given in table \ref{tab:parameters}. \label{fig:steps Biofilm}}
\end{figure}

In Figures \ref{fig:steps Biofilm}, we present a typical sequence of 
snapshots of the bacterial population and the fluid velocity and 
oxygen fields, respectively. In the top row of Fig. \ref{fig:steps Biofilm}, 
each bacterium is represented by a colored dot, with motile bacteria in red, 
the matrix-producer bacteria linked to less than two others bacteria 
in purple, and matrix-producer bacteria linked to at least two other 
bacteria in black. Thus, ``purple'' bacteria are matrix producers 
that are not (yet) firmly integrated in the biofilm structure, 
whereas ``black'' bacteria are part of the connected biofilm tissue.
In the middle and lower row, the maps of the fluid velocity and the
oxygen concentration corresponding to the same times are displayed.

The initial condition of the simulation is a medium at rest, saturated
with oxygen, where 100 planktonic bacteria have been inserted at
random positions. The bacteria move around and divide, and the growth
and aerotaxis process leads in the course of time to an accumulation
of bacteria close to the fluid-air interface as can be seen in Fig.
\ref{fig:7h50}. The density of bacteria close to the interface is
higher than the threshold for the onset of bioconvection, and thus
there is a pronounced fluid motion that can be seen in Fig. \ref{fig:7h50-fluid}.
Due to this convection, the bacterial concentration varies along the
surface, and a small cluster of bacteria that have switched from 
the motile to the matrix-producer phenotype has formed. At
a later stage (fig. \ref{fig:8h10}), the number of bacteria that
have switched is significantly higher, and they have started to bind
together, thus giving birth to a thin biofilm. In addition, the fluid
motion (which is weaker than before, Fig. \ref{fig:8h10-fluid}) has
advected toward the bottom of the container a filament of connected
matrix-producing bacteria that is reminiscent of the filaments observed
in the experiments. Shortly after (Fig. \ref{fig:8h20}), the biofilm
has grown and covers a much larger part of the interface. Finally,
after 11h10 (Fig.~\ref{fig:11h10}), the biofilm covers the whole 
interface and has grown much thicker. { Its surface is, in 
some cases, very irregular and emerges over the level of the water 
surface. This behavior can be attributed to the fact that we do take 
into account neither the gravity force that is exerted on the matrix 
producers that are pushd out of the water by the contact forces, nor 
the capillary forces. The irregular structure of the bulk is probably 
due to the fact that our modeling is purely 2D which makes impossible 
the formation of bicontinuous structures that are more realistic and 
that would allow swimming bacteria  to fill the holes that can be seen 
in the volume of the biofilm.} In addition, there are fewer motile bacteria 
in the medium (because they have switched to matrix producers),
their accumulation at the
surface is less pronounced, and consequently the fluid flow is much weaker.
The oxygen concentration maps during the biofilm formation 
(\ref{fig:7h50-O2}, \ref{fig:8h10-O2}, \ref{fig:8h20-O2}, \ref{fig:11h10-O2}) 
show that there is a strong oxygen gradient close to the
interface and a oxygen-depleted region below. They also indicate that 
the fluid flow induces heterogeneities along the interface (the small
bumps in the oxygen concentration profile are clearly correllated
with convection rolls).  { In fig.\ref{fig:11h10-O2}, one should also 
note that the oxygen concentration map is much more regular than the biofilm
itself, which indicates the averaging effect of diffusion.}

This sequence gives a good illustration of the biofilm formation
process in our model: after a \textit{long} stage (a few hours)
during which bacteria divide and the interplay of oxygen consumption, transport
and bacterial motion leads to an accumulation of bacteria on the
interface, { numerous bacteria switch  from the motile to the matrix-producer state within a
\textit{short} time ($\approx$ 10min -1 hour), which   gives
rise to a thin solid pellicle, the biofilm}. It rapidly covers the
entire interface. Afterwards, the biofilm grows thicker over a few
hours.

Most of the model parameters can be changed over large ranges of 
values without any qualitative change in the scenario outlined
above. However, several parameters have a significant influence 
on both the morphology of the biofilm and on the time between
the beginning of the simulation and the beginning of the biofilm growth,
which we will call {\em nucleation time}. In the following, we
briefly discuss these points, starting with the biofilm morphology
and finishing with the nucleation time.

\subsection{Morphology of the biofilm}

While the rate at which bacteria consume oxygen
or divide has little effect on the final biofilm morphology, the value
of the threshold for the phenotype switch and the mechanical parameters
(elastic constant in the binding force and fluid flow) can dramatically
influence the biofilm morphology. In the following, we briefly describe
these results.

\subsubsection{Effect of the threshold for phenotype switch}

\begin{figure}[H]
\begin{centering}
\includegraphics[height=12cm]{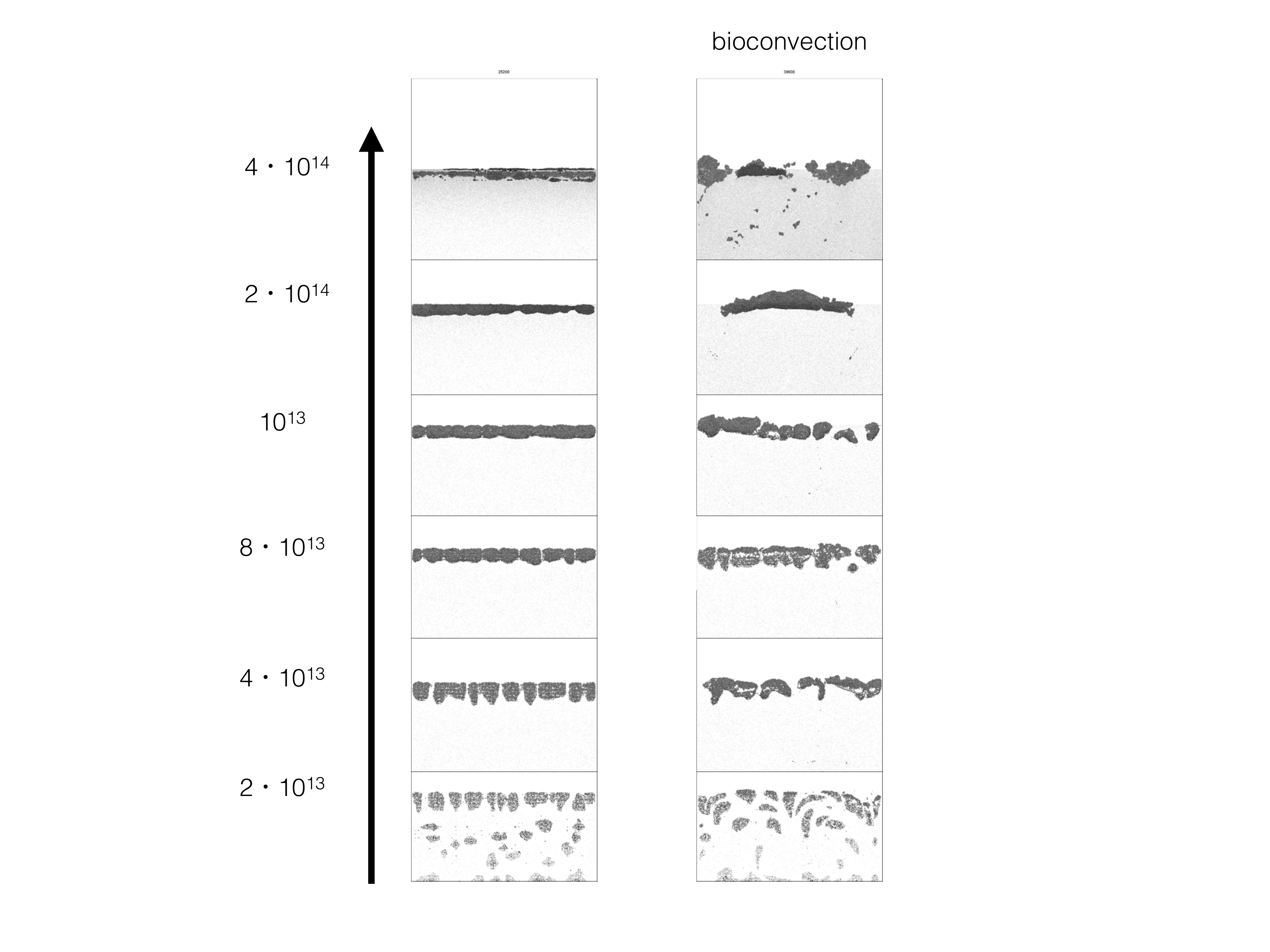} 
\par\end{centering}

\protect\caption{Effect of $n_{ph}$ the threshold of the switch in phenotype on the morphology
of the biofilm, with (right column) and without (left column) bioconvection.
Here, each snapshot is taken when 10000 of bacteria have switched
to the matrix-producer phenotype, and therefore the simulation time
is different for each case. \label{fig:Effect-of-the-threshold-morpholgy}}
\end{figure}

In Fig.~\ref{fig:Effect-of-the-threshold-morpholgy}, we present
a set of simulations that show how the bacterial concentration at
which the phenotype transition occurs ($n_{ph}$) affects the 
morphology of the biofilm, both with and without bioconvection. 
Without bioconvection
(left column), for small values of the threshold, ($2\times10^{13}$
bacteria/m\textsuperscript{3}), biofilm nucleation events are homogeneously
distributed in the whole medium, and the disconnected pieces of biofilm
subsequently grow. For higher values ($4\times10^{13}$ bacteria/m\textsuperscript{3}),
the biofilm is localized close to the interface and consists of chunks
of biofilm that are separated by thin fluid channels. This behaviour
is present (to a certain extent) up to a threshold of $\approx2\times10^{14}$
bacteria/m\textsuperscript{3}. For even higher values, the biofilm is a homogenous
layer at the interface. The presence of bioconvection (right column)
has little effect on the structures. For high values of the threshold, 
the biofilm structures are more disconnected with bioconvection
than without. For smaller values of the threshold, the matrix producers
are organized in structures that are reminiscent of the double convection
roll of the flow.

These observations can be partially understood by taking into account
the different stages of the growth process. When only a few bacteria
are present in the medium, oxygen is supplied by diffusion to the
entire system. When the bacterial concentration exceeds a certain
value, the oxygen in the medium far from the surface is almost completely
consumed, and an oxygen gradient towards the surface  develops (and thus an
oxygen flux towards the bottom). This triggers the migration
of bacteria to the surface. Therefore, both the concentration of the
bacteria at the surface and the bacterial density gradient at the
surface increase with time. If the transition threshold is low, the
transition occurs while the gradients in bacterial concentration are
still relatively low, which explains that nucleation occurs in the
entire system. On the other hand, when the threshold is high, nucleation
occurs only when both the concentration and its gradient are high
at the surface, and therefore biofilm formation occurs only in a thin
layer close to the surface. 

\subsubsection{Effect of the spring constant}

\begin{figure}[H]
\noindent \begin{centering}
\includegraphics[bb=150bp 0bp 600bp 350bp,clip,width=12cm]{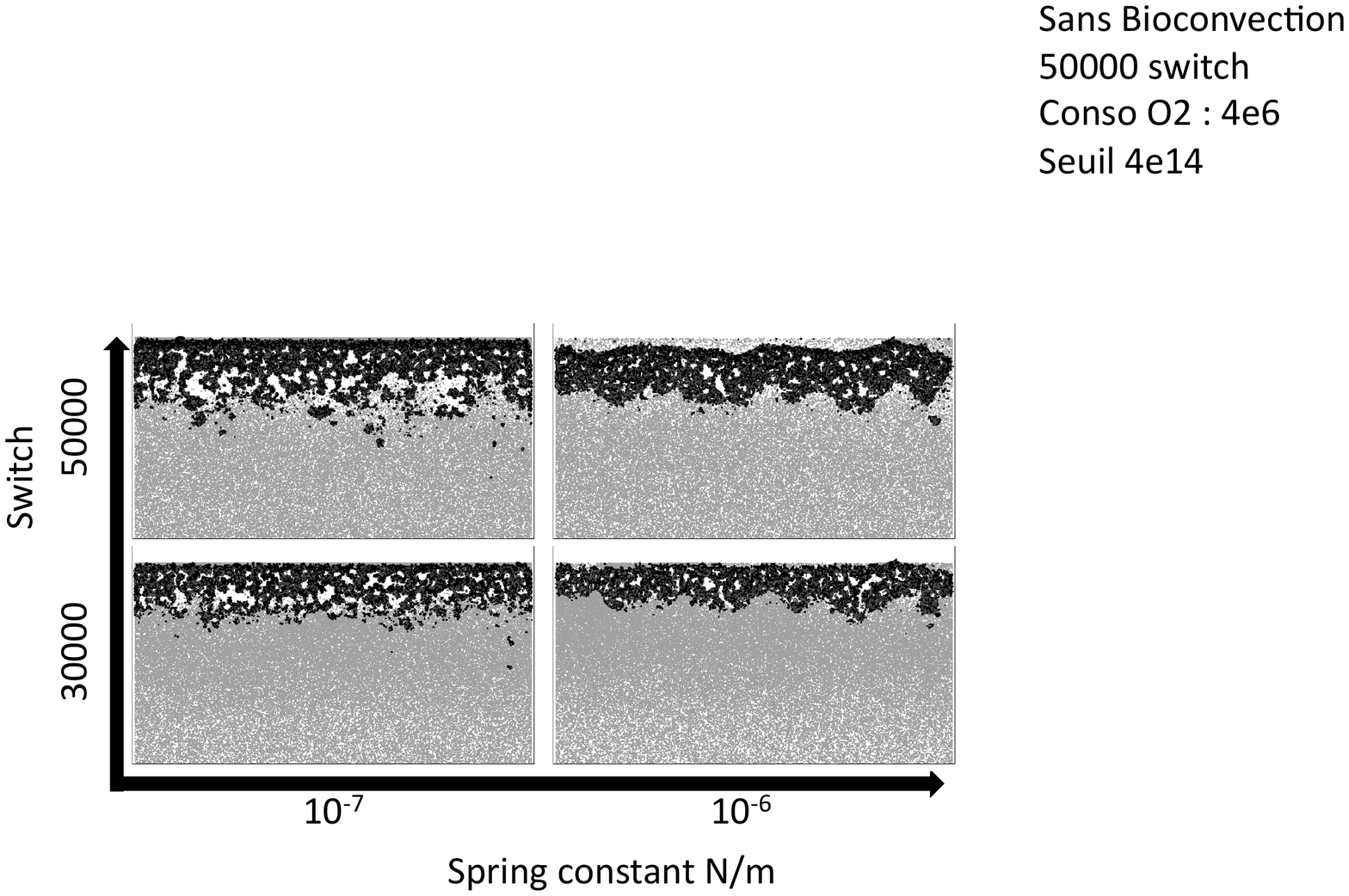} 
\par\end{centering}

\protect\caption{Evolution of the biofilm with different values of spring constant
(without fluid flow). The snapshots show the biofilm structure for
different times when 30000 and 50000 bacteria have switched into matrix
producer phenotype, respectively. \label{fig:spring-constant}}
\end{figure}

There is a double effect upon changing the stiffness of the links between
bacteria. First, this affects the global elastic properties of the biofilm.
Second, having stiffer links implies that they are less likely to
be elongated up to a length at which they will break. This means that
the plasticity of the biofilm will be much smaller and that it will
keep a stronger memory of the growth process than in the case where
the connections between bacteria can easily rearrange to accommodate
external stresses. This can indeed be observed in Fig.~\ref{fig:spring-constant}.
In the case of low elastic constant, the biofilm essentially behaves
like a viscous fluid, and inhomogeneities formed during the initial
stages of growth tend to be smoothed out. In contrast, for higher
elastic constants the growth of the bacterial volume due to matrix
production leads to an accumulation of internal elastic stresses in
the biofilm and, ultimately, to a deformation of the biofilm reminiscent of a buckling phenomenon as can be
seen on the upper right of Fig. \ref{fig:spring-constant}.

\subsection{Nucleation time of the biofilm}

Finally, we consider the effects of the model parameters on the 
time after which the first piece of the biofilm nucleates. Indeed, 
this nucleation time is a quantity that is largely independent of 
the criterion selected to define it, due to the rapid growth of 
the biofilm after its nucleation. Therefore, it can be measured 
with good precision. Here, we consider that nucleation has taken 
place when more than 100 bacteria have changed their phenotype.

\begin{figure}[H]
\begin{centering}
 \subfloat[\label{fig:nucleationTime_avBc}]{\begin{centering}
\includegraphics[width=8cm]{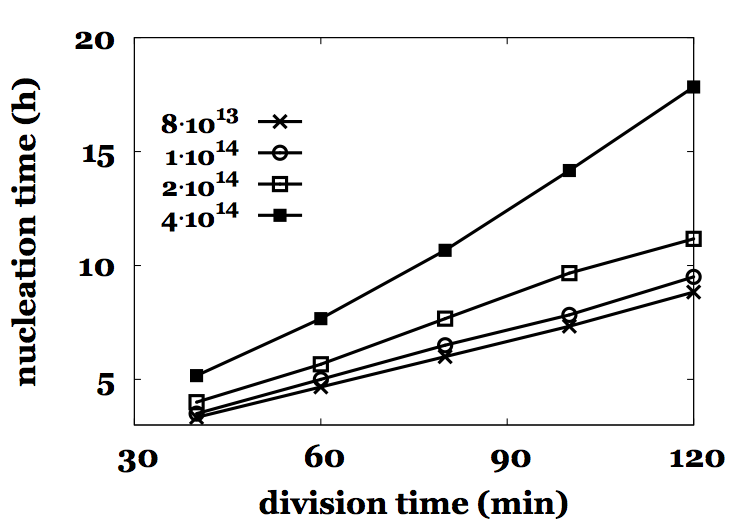}
\par\end{centering}

}\subfloat[\label{fig:nucleationTime_ssBc}]{\begin{centering}
\includegraphics[width=8cm]{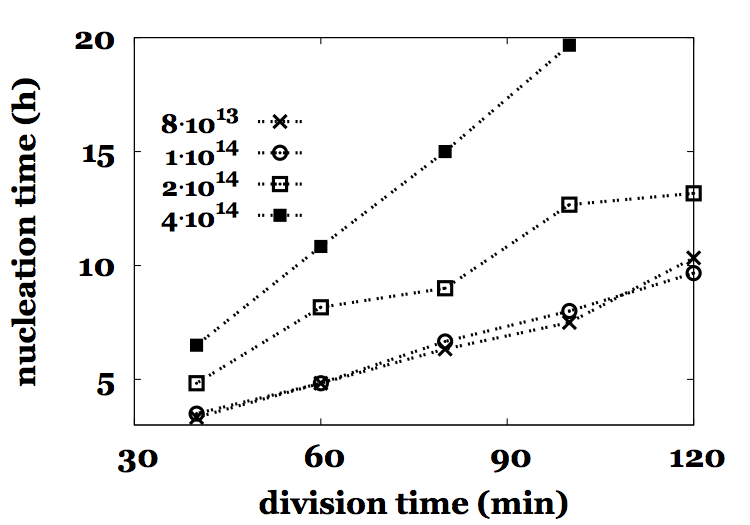}
\par\end{centering}

}
\par\end{centering}

\protect\caption{Nucleation time of the biofilm as a function of the division time $\protect\TimeDiv$ of the bacteria. { The different curves correspond to different values of the phenotype switch threshold  $n_{ph}$}. Figure \ref{fig:nucleationTime_avBc} with convection, figure \ref{fig:nucleationTime_ssBc} without convection. \label{fig:biofilm-apparition-time-division-time}}
\end{figure}

We first consider the role of the division time of the bacteria. In
Fig.~\ref{fig:biofilm-apparition-time-division-time}, the nucleation
time is plotted as a function of the division time of the bacteria
for different values of the phenotype switch threshold, either with
or without flow. In both cases, it is clear that the nucleation time
scales approximately linearly with the division time $\TimeDiv$,
which is to be expected since the division of bacteria is the 
elementary step which governs the population increase which in 
turn triggers the phenotype switch through  quorum sensing. 

\begin{figure}[H]
\begin{centering}
\includegraphics[width=10cm]{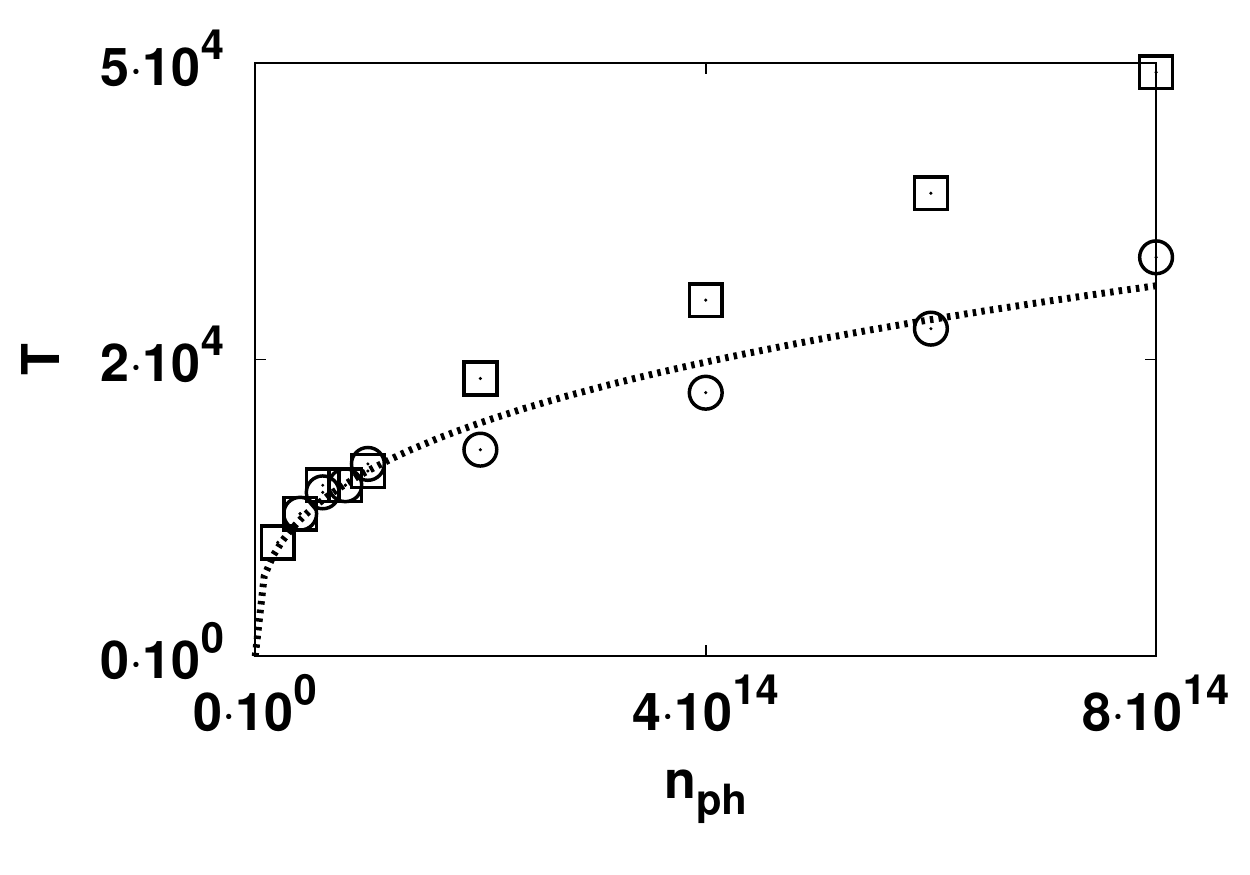} 
\par\end{centering}

\protect\caption{Effect of the threshold in the phenotype switch on the nucleation time of the biofilm. Square dots: simulations with fluid flow. Circle dots: without fluid flow. \label{fig:Effect-of-the-threshold-apparition}}
\end{figure}

We have also studied the nucleation time as a function of the value
of the threshold concentration, both with and without bioconvection.
The results of this study are summarized in Fig.~\ref{fig:Effect-of-the-threshold-apparition}.
In both cases (with and without fluid flow), the nucleation time increases
with the threshold value. For small values of the threshold, the nucleation
times with and without bioconvection are equal (up to numerical uncertainties),
which is expected, since for small values of the threshold,
the biofilm appears before the stratification of the medium is sufficient
to trigger bioconvection. When the threshold is further increased,
biofilm development takes much longer with than without bioconvection.
This is due to the mixing effect of convection, which tends to prevent
bacterial accumulation close to the interface, and therefore to delay
the crossing of the threshold concentration that triggers biofilm
formation.

This result should be compared to experiments of biofilm growth in
which bioconvection can be precisely controlled without significantly 
affecting the environment of the bacteria. To our knowledge, such 
observations have not been reported yet.

\section{Conclusions and perspectives}

\label{sec_conclusion}

We have developed and tested a model of biofilm formation at fluid-air
interfaces. The model combines a continuum description of fluid flow
and oxygen transport with a description of the bacteria as discrete
particles which interact with each other and with their environment.
This model, constructed with a minimal set of hypotheses, reproduces
all the steps of biofilm formation that are observed in the experiments, 
particularly an accumulation of bacteria under the fluid-air interface, 
and bioconvection.

The model relies on several strong simplifying assumptions,
particularly concerning the mechanical interactions of bacteria inside
the biofilm. Since several of the parameters that characterize bacteria
are not well known, quantitative agreement with the experiments cannot
be expected. Nevertheless, with the help of the model we have demonstrated
that bioconvection can significantly influence the time needed 
for biofilms to appear. This is an interesting predicition that 
could be tested in experiements in which bioconvection can be
controlled. Furthermore, we have shown that the biofilm morphology is influenced by the
balance between the accumulation of bacteria close to the oxygen source 
and the quorum-sensing mechanism that triggers the transition from 
motile bacteria to matrix producers.
Therefore, it would be extremely important to have more quantitative
information about the quorum sensing mechanism in order to improve
the model.

Some of the features included in our model are not strictly needed
for the understanding of the initial aggregation and nucleation mechanisms
of biofilms, but will be necessary to model other phenomena observed
in biofilm growth. For instance, after the initial formation of biofilms,
they often develop a characteristic wrinkled morphology during maturation.
This instability has been explained by the accumulation of mechanical
stresses due to the internal growth of the biofilm \cite{Trejo2013}.
The production of a finite volume of matrix in the matrix-producer
state included in our model will naturally lead to the accumulation
of mechanical energy in the biofilm. Nevertheless, a correct description
of the buckling instability would require to develop a coherent treatment
for a non-planar fluid-air interface. More precisely, the effect of
gravity on the emerging bacteria together with a proper description of
capillary effects at the interface should be included. While this is 
probably feasible our model, we rather believe that such phenomena 
should be described in the framework of continuum mechanics. In our 
opinion, promising future lines of research with our model are its
extension to three-dimensional systems so that bicontinuous morphologies 
can appear in the biofilm, and the exploration of changing bacterial 
behavior.

\appendix

\section{Parameters}

Here, we briefly motivate our choices for various model parameters. 
\begin{itemize}
\item Equation (\ref{eq:oxygen}) for the \emph{oxygen consumption} contains two
  parameters : a rate constant $\gamma$ and a Michaelis constant $K$. The oxygen
  consumption rate of \textit{B. subtilis} is $10^{6}$~molecules$/$s$/$bacterium
  in a saturated culture \cite{Cisneros2008these}. It was shown by Martin for
  \textit{Escherichia coli} that this rate can vary by one order of magnitude
  depending on the growth phase of the bacteria \cite{Martin1932}, with
  saturated culture corresponding to the minimal oxygen uptake. We suppose that
  similar variations can occur for \textit{B. subtilis}, and thus $\gamma$
  varies in the range of $10^{6} -10^{7}$~molecules$/$s$/$bacterium. We have
  used a Michaelis-Menten law to cut off the oxygen consumption at low
  concentrations; the corresponding Michaelis constant $K$ is unknown. Since
  observations on\emph{ E. coli} indicate that oxygen is almost completely
  depleted in concentrated cultures \cite{Douarche2009_PRL}, we choose a very
  small value of $K$ compared to the initial oxygen concentration. 

\item The \emph{integration time constants} of the oxygen memory,
  $\TimeMemoryShort$ and $\TimeMemoryLong$ in equations (\ref{eq:shortMeme}) and
  (\ref{eq:longMeme}) are the typical time intervals over which the oxygen
  concentration is averaged in the internal variables $\MemoryShort$ and
  $\MemoryLong$, respectively. Experimental observations have shown that
  \textit{B. subtilis} is able to detect quickly (less than $1\, s$) a sudden
  variation in oxygen concentration, but adapts to the new average level of
  oxygen within several seconds \cite{Wong1995}. According to these results we
  take $\TimeMemoryShort=0.1\, s$ and $\TimeMemoryLong=10\, s$. 

\item The \emph{radius} $\Radius$ is used for the calculation of the bacterial
  velocity in Eq.~(\ref{eq:bacterial velocity}). We choose a \emph{reference
  radius} $\Radius_{0}=5\,\mu$m (the typical length of \textit{B. subtilis} when
  it is in the motile phenotype \cite{Cisneros2008these}). For simplicity, we
  keep the friction coefficient $6\pi\eta r$ constant in Eq.~(\ref{eq:bacterial
  velocity}) by always using $\Radius=\Radius_{0}$. However, in order to
  properly calculate the forces between bacteria,
  Eq.~(\ref{eq:force_elastique}), we take into account the growth of the
  bacterial body size with time through the value of
  $\Radius_{ij}=\Radius_{i}+\Radius_{j}$. The maximum radius of motile bacteria,
  $\RadiusDiv$ is chosen as $\RadiusDiv=\sqrt[3]{2}r_0\backsimeq6,3\,\mu m$,
  which corresponds to a volume twice larger than the reference volume. 

\item The \emph{mass density} of bacteria $\rho_{b}$ is needed in
  Eq.~(\ref{eq:density}) to evaluate the local fluid density for use in the
  Navier-Stokes equation (\ref{eq:NS}). As already mentioned, several values for
  this density are quoted in the literature. We take for our simulations a
  density that is $3\%$ larger than the one of the medium. However, the mature
  biofilm usually floats on the water, which means that it must also contain
  some components that are lighter than water. The mass density of the
  extracellular matrix is actually unknown. To take these observations into
  account in a simple manner, we use the bacterial ``reference volume''
  $4\pi\Radius_{0}^{3}/3$ for each motile bacterium in the calculation of the
  total bacterial volume $V_{b}$ in a coarse-grained cell, whereas matrix
  producers do not contribute. 

\item To determine the \emph{division time} of \textit{B. subtilis} during
  growth in biofilm conditions, we measure the evolution of the bacterial
  concentration in the medium over time. The measured division time is around
  1h, and we take for the simulations $\TimeDiv=70$ min.

\item The \emph{propulsion velocity} is in the range of $10\,\mu m.s^{-1}$ to
  $30\,\mu m.s^{-1}$ and slightly depends on the local oxygen concentration
  \cite{Sokolov2012}. We have taken a constant $\Speedbacteria_{0}=20\,\mu
  m.s^{-1}$ for simplicity. 

\item The rate of switching from the motile to the matrix-producer phenotype,
  Eq.~(\ref{eq:proba threshold}), contains two constants: the \emph{ quorum
  sensing} threshold $\BacterialCon_{ph}$ and the rate $1/\TimePheno$. We
  observe in the experiments that the bacterial concentration in the medium at
  the time of the beginning of the biofilm formation is around $10^{13}$
  bacteria$/$m$^{3}$. The threshold for the change in phenotype must then be
  higher than this value, because at the water-air interface the bacterial
  concentration is higher than in the bulk. We explore various values of this
  parameter in the simulations. The switching time $\TimePheno$ sets the rate of
  switching when concentration threshold is exceeded. We suppose that the
  transition happens quickly and take $\TimePheno=\TimeStep$.\footnote{ We also performed
  simulations with larger values of  $\TimePheno=\TimeStep$ in the range of 1 to 10
  minutes without noticing any significant effect. }
\end{itemize}


\end{document}